\documentclass[11pt,pre,superscriptaddress,showkeys,byrevtex,bibnotes]{revtex4}
\usepackage{times}

\newtheorem{defn}{Definition}[section]

\newtheorem{corollary}[defn]{Corollary}
\newtheorem{example}[defn]{Example}

\newtheorem{prop}[defn]{Proposition}

\newtheorem{remark}[defn]{Remark}
\newtheorem{thm}[defn]{Theorem}
\newtheorem{theorem}[defn]{Theorem}
\newcommand{\be}{\begin{equation}}
\newcommand{\ee}{\end{equation}}
\newcommand{\bea}{\begin{eqnarray}}
\newcommand{\eea}{\end{eqnarray}}
\newcommand{\beas}{\begin{eqnarray*}}
\newcommand{\eeas}{\end{eqnarray*}}
\newcommand{\brmk}{\begin{remark}\per\begin{em}}
\newcommand{\ermk}{\end{em}\end{remark}}
\newcommand{\goto}{\rightarrow}
\newcommand{\ink}{\rule{.5\baselineskip}{.55\baselineskip}}
\newcommand{\noi}{\noindent}
\newcommand{\lan}{\langle}
\newcommand{\ran}{\rangle}

\newcommand{\skp}{\vspace{\baselineskip}}

\newcommand{\per}{\hspace{-.072in}{\bf .  }}

\newcommand{\R}{I\!\!R}
\renewcommand{\r}{I\!\!R}

\newcommand{\N}{{I\!\!N}}
\newcommand{\ZZ}{{Z\!\!\!Z}}
\newcommand{\rsigma}{I\!\!R^{\sigma}}

\newcommand{\e}{{\cal E}}
\newcommand{\eu}{{\cal E}^u}
\newcommand{\ebeta}{{\cal E}_\beta}
\newcommand{\egbeta}{{\cal E}(g)_\beta}
\newcommand{\egammabeta}{{\cal E}(\gamma)_\beta}
\newcommand{\X}{{\cal X}}

\newcommand{\thi}{\tilde{H}}

\newcommand{\mboxdomf}{\mbox{dom} \, f}
\newcommand{\mboxemdomf}{\mbox{{\em dom}} \, f}
\newcommand{\mboxridomf}{\mbox{ri(dom} \, f)}
\newcommand{\mboxemridomf}{\mbox{{\em ri(dom}} \, f)}
\newcommand{\mboxintdomf}{\mbox{int(dom} \, f)}

\newcommand{\mboxdoms}{\mbox{dom} \, s}
\newcommand{\mboxemdoms}{\mbox{{\em dom}} \, s}
\newcommand{\mboxridoms}{\mbox{ri(dom} \, s)}
\newcommand{\mboxemridoms}{\mbox{{\em ri(dom}} \, s)}
\newcommand{\mboxintdoms}{\mbox{int(dom} \, s)}
\newcommand{\mboxemintdoms}{\mbox{{\em int(dom}} \, s)}
\newcommand{\bexa}{\begin{example}\per\begin{em}}
\newcommand{\eexa}{\end{em}\end{example}}
\newcounter{bean}
\newcommand{\benuma}{\setlength{\labelwidth}{.25in}
\begin{list}%
{(\alph{bean})}{\usecounter{bean}}}
\newcommand{\eenuma}{\end{list}}
\def\theequation{\thesection.\arabic{equation}}
\def\theequation{\arabic{section}.\arabic{equation}}
\def\thedefn{\arabic{section}.\arabic{defn}}

\begin{document}

\title{The Generalized Canonical Ensemble
and Its Universal Equivalence
with the Microcanonical Ensemble}

\author{M. Costeniuc}%
\email{costeniuc@math.umass.edu}
\affiliation{Department of Mathematics and Statistics, University of Massachusetts,
Amherst, Massachusetts, USA 01003} %

\author{R. S. Ellis}%
\email{rsellis@math.umass.edu}%
\affiliation{Department of Mathematics and Statistics, University of Massachusetts,
Amherst, Massachusetts, USA 01003} %

\author{H. Touchette}%
\email{htouchet@alum.mit.edu}%
\affiliation{School of Mathematical Sciences, Queen Mary, University of London,
London, UK E1 4NS}%

\author{B. Turkington}%
\email{turk@math.umass.edu}
\affiliation{Department of Mathematics and Statistics, University of Massachusetts,
Amherst, Massachusetts, USA 01003} %

\date{\today}%

\begin{abstract}
Microcanonical equilibrium macrostates are characterized as the solutions of a
constrained minimization problem, while canonical equilibrium macrostates
are characterized as the solutions of a related, unconstrained minimization problem.
In the paper \cite{EHT1} by Ellis, Haven, and Turkington, the
problem of ensemble equivalence was completely solved at two separate,
but related levels: the level of equilibrium macrostates,
which focuses on relationships between the corresponding sets
of equilibrium macrostates, and the thermodynamic level, which
focuses on when the microcanonical
entropy $s$ can be expressed as the Legendre-Fenchel transform
of the canonical free energy.
A neat but not
quite precise statement of the main result in \cite{EHT1}
is that the microcanonical and canonical ensembles
are equivalent at the level of equilibrium macrostates
if and only if they are equivalent
at the thermodynamic level, which is the case if and only if the microcanonical
entropy $s$ is concave.

The present paper extends the results in \cite{EHT1} significantly by
addressing the following motivational question.
Given that the microcanonical ensemble is not equivalent
with the canonical ensemble, is it possible to replace the canonical ensemble
with a generalized canonical ensemble that is equivalent with the
microcanonical ensemble?  The generalized canonical ensemble
that we consider is obtained from
the standard canonical ensemble by adding an exponential factor
involving a continuous function $g$ of the Hamiltonian.
The special case
in which $g$ is quadratic plays a central role
in the theory, giving rise to a generalized canonical
ensemble known in the literature as the Gaussian ensemble.

As in \cite{EHT1}, we analyze
the equivalence of the two ensembles at both the level of equilibrium macrostates
and the thermodynamic level.  A neat but not
quite precise statement of the main result in the present paper
is that the microcanonical and generalized canonical ensembles
are equivalent at the level of equilibrium macrostates
if and only if they are equivalent
at the thermodynamic level, which is the case if and only if
the generalized microcanonical entropy $s-g$ is concave.
The considerable freedom that one has in choosing $g$ has the important
consequence that even
when the microcanonical and standard canonical ensembles are not equivalent,
one can often find $g$ with the property that the microcanonical
and generalized canonical ensembles satisfy a strong form of equivalence which
we call universal equivalence.  For example, if the microcanonical entropy
is $C^2$, then universal equivalence of ensembles holds with $g$ taken
from a class of quadratic functions.  This use
of functions $g$ to obtain ensemble equivalence is a counterpart to
the use of penalty functions and augmented Lagrangians in global optimization.
\end{abstract}


\maketitle

\section{Introduction}
\setcounter{equation}{0}
\label{section:intro}

The problem of ensemble equivalence is a fundamental one lying at the foundations
of equilibrium statistical mechanics.  When formulated in mathematical
terms, it is apparent that this problem also addresses a fundamental
issue in global optimization.  Given a constrained minimization problem, 
under what conditions does there exist a related, unconstrained minimization problem having the same 
minimum points?

In order to explain the connection between ensemble equivalence
and global optimization and in order to outline the contributions of this paper, we
introduce some notation.  Let ${\cal X}$ be a space,
$I$ a function mapping ${\cal X}$ into $[0,\infty]$, and $\tilde{H}$ a function
mapping ${\cal X}$ into $\rsigma$, where $\sigma$ is a positive integer.  
For $u \in \rsigma$ we consider the following constrained minimization problem:
\be
\label{eqn:constrained}
\mbox{minimize } I(x) \mbox{ over } x \in {\cal X}
\mbox{ subject to the contraint } \tilde{H}(x) = u.
\ee
A partial answer to the 
question posed at the end of the first paragraph can be found by
introducing the following related, unconstrained minimization problem
for $\beta \in \rsigma$:
\be
\label{eqn:unconstrained}
\mbox{minimize } I(x) + \lan \beta,\tilde{H}(x) \ran 
\mbox{ over } x \in \X,
\ee
where $\lan \cdot, \cdot \ran$ denotes the Euclidean inner product on 
$\rsigma$.   The theory of Lagrange multipliers outlines suitable 
conditions under which the solutions of the constrained problem
(\ref{eqn:constrained}) lie among the 
critical points of $I + \lan \beta,\tilde{H} \ran$.  However, it does not give, 
as we will do in Theorems \ref{thm:equiv} and \ref{thm:main}, necessary
and sufficient conditions for the solutions of (\ref{eqn:constrained}) to coincide with
the solutions of the unconstrained minimization problem (\ref{eqn:unconstrained}).
By giving such necessary and sufficient conditions, we make contact
with the duality theory of global optimization and the method of
augmented Lagrangians \cite[\S2.2]{Bertsekas}, \cite[\S6.4]{Minoux}.
In the context of global optimization the primal function and the dual
function play the same roles that the (generalized) microcanonical
entropy and the (generalized) canonical free energy play in
statistical mechanics.  Similarly, the replacement of the Lagrangian
by the augmented Lagrangian in global optimization is paralleled by
our replacement of the canonical ensemble by the generalized canonical
ensemble.

The two minimization problems (\ref{eqn:constrained})
and (\ref{eqn:unconstrained}) arise in a natural way in 
the context of equilibrium statistical mechanics \cite{EHT1}, where
in the case $\sigma = 1$, 
$u$ denotes the mean energy and $\beta$ the inverse temperature.
We define $\e^u$
and $\e_\beta$ to be the respective sets of points solving the constrained
problem (\ref{eqn:constrained}) and the unconstrained
problem (\ref{eqn:unconstrained}); i.e.,
\be
\label{eqn:euintro}
\eu = \{x \in \X : I(x) \mbox{ is minimized subject to } \thi(x) = u\}
\ee
and
\be
\label{eqn:ebetaintro}
\ebeta = \{x \in \X : I(x) + \lan \beta,\thi(x) \ran \mbox{ is minimized}\}.
\ee
For a given statistical mechanical model $\X$ represents the set of all
possible equilibrium macrostates.  As we will outline in Section 2, 
the theory of large deviations allows one to identify $\eu$ as
the subset of $\X$ consisting of equilibrium macrostates for the microcanonical ensemble
and $\e_\beta$ as the subset consisting
of equilibrium macrostates for the canonical ensemble.

Defined by conditioning the Hamiltonian to have a fixed value, 
the microcanonical ensemble expresses the conservation 
of physical quantities such as the energy and is the more fundamental
of the two ensembles.  Among other reasons, the 
canonical ensemble was introduced by Gibbs \cite{Gibbs} in the hope that 
in the limit $n \goto \infty$ the two
ensembles are equivalent; i.e., 
all asymptotic properties of the model obtained via the microcanonical ensemble
could be realized as asymptotic properties obtained via the canonical ensemble.  
However, as numerous studies discussed near the end of
this introduction have shown, in general this is not the case.  
There are many examples of statistical mechanical models for which
nonequivalence of ensembles holds over a wide range of model
parameters and for which physically interesting microcanonical
equilibria are often omitted by the canonical ensemble.

The paper \cite{EHT1} investigates this question in detail,
analyzing equivalence of ensembles in terms of 
relationships between $\eu$ and $\ebeta$.  In turn, these relationships
are expressed in terms of support 
and concavity properties of the microcanonical entropy
\[
\label{eqn:intros}
s(u) = - \inf \{I(x) : x \in \X, \tilde{H}(x) = u\}.
\]
The main results in \cite{EHT1} are summarized in Theorem \ref{thm:equiv},
which we now discuss under the simplifying assumption
that $\mboxdoms$ is an open subset of $\rsigma$.

We focus on $u \in \mboxdoms$.  Part (a) of Theorem \ref{thm:equiv} 
states that if $s$ has a strictly supporting hyperplane 
at $u$, then full equivalence of ensembles holds in the sense that there
exists a $\beta$ such that $\eu = \ebeta$.  In particular,
if $\mboxdoms$ is convex and open and $s$ is strictly concave on $\mboxdoms$, then 
$s$ has a strictly supporting hyperplane at all $u$ 
[Thm.\ \ref{thm:strictlyconcave}(a)] and thus
full equivalence of ensembles holds at all $u$. 
In this case we say that the microcanonical
and canonical ensembles are {\bf universally equivalent}.  

The most surprising result, given in part (c), 
is that if $s$ does not have a supporting hyperplane at $u$,
then nonequivalence of ensembles holds in the strong sense that
$\eu \cap \ebeta = \emptyset$ for all $\beta \in \rsigma$.  That is,
if $s$ does not have a supporting hyperplane at $u$ --- equivalently, 
if $s$ is not concave at $u$ --- then
microcanonical equilibrium macrostates cannot be realized canonically.
This is to be contrasted with part (d), which states
that for any $x \in \ebeta$ there exists $u$ such that
$x \in \eu$; i.e., canonical equilibrium macrostates can always be realized
microcanonically.  Thus of the two ensembles the microcanonical is the richer.

The starting point of the present paper is the following motivational question
suggested by Theorem \ref{thm:equiv}. Given that the microcanonical ensemble is not equivalent with the
canonical ensemble on a subset of values of $u$, is it possible to replace the canonical ensemble
with a generalized canonical ensemble that is univerally equivalent with the 
microcanonical ensemble; i.e., fully equivalent at all $u$?

The generalized canonical ensemble that we consider 
is a natural perturbation of the standard canonical ensemble, obtained from 
it by adding an exponential factor
involving a continuous function $g$ of the Hamiltonian.
The special case
in which $g$ is quadratic plays a central role 
in the theory, giving rise to a generalized canonical
ensemble known in the literature as the Gaussian ensemble 
\cite{CH1,CH2,Heth1987,Stump1987,JPV,Stump21987}.  As these papers discuss, an important feature 
of Gaussian ensembles is that they 
allow one to account for ensemble-dependent effects in finite systems.
Although not referred to by name, the Gaussian ensemble
also plays a key role in \cite{KieLeb}, where it is used to address
equivalence-of-ensemble questions for a point-vortex model of fluid
turbulence.  

Let us focus on the case of quadratic $g$ because it illustrates nicely
why the answer to the motivational question is yes in a wide variety of circumstances.  
In order to simplify
the notation, we work with $u = 0$ and the 
corresponding set ${\cal E}^0$ of equilibrium macrostates.  
We denote by $\|\cdot\|$ the Euclidean norm on $\rsigma$ and consider
the Gaussian ensemble defined in (\ref{eqn:gencanon})
with $g(u) = \gamma \|u\|^2$ for $\gamma \geq 0$.
As we will outline in Section 2, the theory of large deviations
allows one to identify the subset of $\X$ consisting of equilibrium 
macrostates for the Gaussian ensemble with the set
\be
\label{eqn:egammabeta}
\egammabeta = \left\{x \in \X : I(x) + \lan \beta,\thi(x) \ran + \gamma \|\thi(x)\|^2
\mbox{ is minimized}\right\}.
\ee

$\egammabeta$ can be viewed as an approximation to the
set ${\cal E}^0$ of equilibrium macrostates for the microcanonical ensemble.
This follows from the calculation 
\beas
\lefteqn{
\left\{x \in \X : \lim_{\gamma \goto \infty} \left(I(x) + \lan \beta,\thi(x) \ran + 
\gamma \|\thi(x)\|^2\right) \mbox{ is minimized}\right\}} \\ 
&& = \{x \in \X : I(x) \mbox{ is minimized subject to } \thi(x) = 0\} = {\cal E}^0.
\eeas
This observation makes it plausible that there exist a $\beta$
and a sufficiently large $\gamma$ such that ${\cal E}^0$ equals
$\egammabeta$; i.e., the microcanonical ensemble and the Gaussian ensemble
are fully equivalent.  As we will see, under suitable hypotheses this and much more are true.

Our results apply to a much wider class of generalized canonical ensembles,
of which the Gaussian ensemble is a special case.  Given a continuous
function $g$ mapping $\rsigma$ into $\R$, 
the associated set of equilibrium macrostates is defined as
\[
\egbeta = \{x \in \X : I(x) + \lan \beta,\thi(x) \ran + g(\thi(x))
\mbox{ is minimized}\}.
\]
This set reduces to (\ref{eqn:egammabeta}) when $g(u) = \gamma \|u\|^2$.

The utility of the generalized canonical ensemble rests on the simplicity with
which the function $g$ defining this ensemble
enters the formulation of ensemble equivalence.  
Essentially all the results in \cite{EHT1}
concerning ensemble equivalence, including
Theorem \ref{thm:equiv}, generalize
to the setting of the generalized canonical ensemble by replacing
the microcanonical entropy $s$ by the generalized microcanonical entropy $s-g$.  
The generalization of Theorem \ref{thm:equiv}
is stated in Theorem \ref{thm:main},
which gives all possible relationships
between the set $\eu$ of equilibrium macrostates for the microcanonical ensemble
and the set $\egbeta$ of equilibrium macrostates for
the generalized canonical ensemble.  These relationships
are expressed in terms of support and concavity properties of $s-g$.
The proof of Theorem \ref{thm:main} shows how easily it follows 
from Theorem \ref{thm:equiv}, 
in which all equivalence and nonequivalence relationships between
$\eu$ and $\ebeta$ are expressed in terms of support and concavity properties
of $s$.

For the purpose of applications the most important consequence of Theorem \ref{thm:main}
is given in part (a), which we now discuss
under the simplifying assumption that $\mboxdoms$ is an open subset of $\rsigma$.
We focus on $u \in \mboxdoms$.   
Part (a) states that if $s - g$ has a strictly supporting hyperplane
at $u$, then full equivalence of ensembles holds in the sense that
there exists a $\beta$ such that $\eu = \egbeta$.  In particular,
if $\mboxdoms$ is convex and open and 
if $s-g$ is strictly concave on $\mboxdoms$, then $s-g$
has a strictly supporting hyperplane at all $u$ 
[Thm.\ \ref{thm:sgstrictlyconcave}(a)] and thus full equivalence
of ensembles holds at all $u$.  In this case we say that the microcanonical and
generalized canonical ensembles are {\bf universally equivalent}.

The only requirement on the function $g$ defining
the generalized canonical ensemble is that $g$ is continuous.
The considerable freedom that one has in choosing $g$
makes it possible to define a generalized canonical ensemble
that is universally equivalent with the 
microcanonical ensemble when the microcanonical
and standard canonical ensembles are not equivalent on 
a subset of values of $u$.  In Theorems 
\ref{thm:applysigma1}--\ref{thm:apply2}
several examples of universal equivalence
are derived under natural smoothness and boundedness conditions on $s$,
while Theorem \ref{thm:localapply} derives
a weaker form of universal equivalence under other conditions.
In the first, second, and fourth of these theorems $g$ is taken from a set of 
quadratic functions, and the associated ensembles are Gaussian.

Theorem \ref{thm:applysigma1}, which applies when the dimension $\sigma = 1$, 
is particularly useful.  It shows that if $s$ is $C^2$ and 
$s''$ is bounded above on the interior of $\mboxdoms$, 
then for any 
\[
\gamma > \textstyle \frac{1}{2} \cdot \displaystyle\sup_{x \in \mbox{\scriptsize int(dom} \, s)} s''(x),
\]
$s(u) - \gamma u^2$ is strictly concave on $\mboxdoms$. 
By part (b) of Theorem \ref{thm:sgstrictlyconcave} and
part (a) of Theorem \ref{thm:main}, it follows that the 
microcanonical ensemble and the Gaussian ensemble defined in terms of $\gamma$
are universally equivalent.
The strict concavity of $s(u) - \gamma u^2$ also implies that the generalized canonical free energy
is differentiable on $\r$ [Thm. \ref{thm:properties}(c)], a condition
guaranteeing the absence of a discontinuous, first-order phase transition 
with respect to the Gaussian ensemble.
Theorem \ref{thm:apply} is the analogue of Theorem \ref{thm:applysigma1}
that treats arbitrary dimension $\sigma \geq 2$.  Again, we prove that for 
all sufficiently large $\gamma$, the microcanonical ensemble and the 
Gaussian ensemble defined in terms of $\gamma$
are universally equivalent.  These two theorems are particularly 
satisfying because they make rigorous the intuition
underlying the introduction of the Gaussian ensemble: because
it approximates the microcanonical ensemble in the limit $\gamma \goto \infty$,
universal ensemble equivalence should hold for all sufficiently large $\gamma$.

The criterion in Theorem \ref{thm:applysigma1} that $s''$
is bounded above on the interior of $\mboxdoms$ is essentially
optimal for the existence of a fixed
quadratic function $g$ guaranteeing the strict concavity of $s - g$ on $\mboxdoms$.
The situation
in which $s''(u) \goto \infty$ as $u$ approaches a boundary point 
can often be handled by Theorem \ref{thm:localapply}, which is a local version of
Theorem \ref{thm:applysigma1}.   

Besides studying ensemble equivalence 
at the level of equilibrium macrostates, one can also analyze it at the thermodynamic level.  
This level focuses on Legendre-Fenchel-transform relationships involving 
the basic thermodynamic functions in the three ensembles:
the microcanonical entropy $s(u)$, on the one hand,
and the canonical free energy and generalized canonical free energy, on the other.
The analysis is carried out in Section \ref{section:thermo}, where we
also relate ensemble equivalence at the two levels. 
A neat but not quite precise statement of the main result proved in that section 
is that the microcanonical ensemble and the canonical ensemble
(resp., generalized canonical ensemble)
are equivalent at the level of equilibrium macrostates
if and only if they are equivalent
at the thermodynamic level, which is the case if and only $s$ (resp., $s-g$) is concave.

One of the seeds out of which the present paper germinated is the paper
\cite{EHT2}, in which we study the equivalence of the microcanonical
and canonical ensembles for statistical equilibrium models of coherent
structures in two-dimensional and quasi-geostrophic turbulence.
Numerical computations demonstrate that nonequivalence of ensembles occurs
over a wide range of model parameters and that physically interesting
microcanonical equilibria are often omitted by
the canonical ensemble.  In addition, in 
Section 5 of \cite{EHT2}, we establish the nonlinear stability of the steady
mean flows corresponding to microcanonical equilibria 
via a new Lyapunov argument.  The associated stability theorem refines the
well-known Arnold stability theorems, which do not apply when the
microcanonical and canonical ensembles are not equivalent.  The Lyapunov
functional appearing in this new stability 
theorem is defined in terms of a generalized thermodynamic potential similar in form to 
\[
I(x) + \lan \beta,\thi(x) \ran + \gamma \|\thi(x)\|^2,
\]
the minimum points of which define the set of equilibrium macrostates
for the Gaussian ensemble [see (\ref{eqn:egammabeta})].  
Such Lyapunov functionals arise
in the study of constrained optimization problems, where they are known as
augmented Lagrangians \cite{Bertsekas,Minoux}.  

Another seed out of which the present paper germinated is the work of
Hetherington and coworkers \cite{CH1,CH2,Heth1987,Stump1987,Stump21987} on the
Gaussian ensemble.  Reference \cite{Heth1987} is the first paper that defined
the Gaussian ensemble as a modification of the canonical ensemble
in which the standard exponential
Boltzmann term involving the energy is augmented by an additional
term involving the square of the energy.  As shown in 
\cite{CH1,CH2,Stump1987,Stump21987}, such a modified canonical 
ensemble arises when a sample system is
in contact with a finite heat reservoir.  From this point of
view, the Gaussian ensemble can be viewed as an intermediate ensemble
between the microcanonical, whose definition involves no reservoir,
and the canonical ensemble, which is defined in terms of an infinite reservoir. The Gaussian
ensemble is used in \cite{CH1,CH2,Stump1987,Stump21987} to study
microcanonical-canonical discrepancies in finite-size systems; such discrepancies are
generally present near first-order phase transitions.

Gaussian ensembles are also considered in [32] and more or less implicitly
in [33]. Reference [32] is a theoretical study of the Gaussian ensemble
which derives it from the maximum entropy principle and studies its
stability properties. The second paper [33] uses some mathematical
methods that are reminiscent of the Gaussian ensemble to study a
point-vertex model of fluid turbulence.  By sending $\gamma \goto \infty$
after the fluid limit $n \goto \infty$, the authors recover the special class of 
nonlinear, stationary Euler flows that is expected from the microcanonical
ensemble.  Their use of Gaussian ensembles improves previous studies in which 
either the logarithmic singularities of the Hamiltonian must be 
regularized or equivalence of ensembles must be assumed.  As they point out,
the latter is not a satisfactory assumption because the ensembles
are nonequivalent in certain geometries in which conditionally stable configurations
exist in the microcanonical ensemble but not in the canonical ensemble.  Their
paper motivated in part the analysis of ensemble equivalence in the present
paper, which focuses on generalized canonical ensembles with a fixed function
$g$ and, as a special case, Gaussian ensembles in which $\gamma$ is fixed and is not
sent to $\infty$.

In addition to the connections with \cite{CH1,CH2,JPV,KieLeb}, 
the present paper also builds on the wide literature concerning
equivalence of ensembles in statistical mechanics.  An overview
of this literature is given in the introduction of \cite{LewPfiSul2}.  
A number of papers on this topic, including 
\cite{DeuStrZes,EHT1,EyiSpo,Geo,LewPfiSul1,LewPfiSul2,RoeZes}, investigate
equivalence of ensembles using the theory of large deviations.  
In \cite[\S7]{LewPfiSul1} and \cite[\S7.3]{LewPfiSul2}   
there is a discussion of  nonequivalence of ensembles for the 
simplest mean-field model in statistical mechanics; namely, the
Curie-Weiss model of a ferromagnet.  
However, despite the mathematical
sophistication of these and other studies, none of them except
for our paper \cite{EHT1} explicitly addresses
the general issue of the nonequivalence of ensembles, which seems
to be the typical behavior for a wide class of models arising
in various areas of statistical mechanics.

Nonequivalence of ensembles
at the thermodynamic level has been observed in a number of long-range, mean-field spin models,
including the Hamiltonian mean-field model \cite{DauLatRapRufTor,LatRapTsa2},
the mean-field X-Y model \cite{DHR}, 
and the mean-field Blume-Emery-Griffith model \cite{BMR1,BMR2}.  
In \cite{ETT} ensemble nonequivalence
for the mean field Blume-Emery-Griffiths model was demonstrated
to hold also at the level of equilibrium macrostates via numerical
computations.  For a mean-field version of the 
Potts model called the Curie-Weiss-Potts model, equivalence and nonequivalence of ensembles at the 
level of equilibrium macrostates is analyzed in detail in \cite{CosEllTou1,CosEllTou2}.
Ensemble nonequivalence has also been observed in models
of turbulence \cite{CLMP,EHT2,EyiSpo,KieLeb,RobSom}, models of plasmas
\cite{KieNeu2,SmiOne}, gravitational systems \cite{Gross1,HerThi,LynBelWoo,Thi2},
and a model of the Lennard-Jones gas \cite{BorTsa}.
Many of these models can also be
analyzed by the methods of \cite{EHT1} and the present paper.  A detailed
discussion of ensemble nonequivalence for models of turbulence is given
in \cite[\S1.4]{EHT1}.

The study of ensemble equivalence at the level of equilibrium macrostates
involves relationships among the sets $\eu$, $\ebeta$, and $\egbeta$ of
equilibrium macrostates for the three ensembles.  These sets are subsets
of $\X$, which in many cases, including short-range spin models
and models of turbulence, is an infinite dimensional space.  The most
important discovery 
in our work on this topic is that all relationships among these possibly
infinite dimensional sets are completely determined by support and
concavity properties of the finite-dimensional, and in many applications,
one-dimensional functions $s$ and $s-g$.  The main 
tools for analyzing ensemble equivalence are the theory of large 
deviations and the theory of concave functions, both of which exhibit
an analogous conceptual structure.  On the one hand,
the two theories provide powerful, investigative methodologies in which formal manipulations
or geometric intuition can lead one to the correct answer.  On the other hand,
both theories are fraught with numerous technicalities which, if emphasized,
can obscure the big picture.  In the present paper we emphasize
the big picture by relegating a number of technicalities to the appendix.
The reference \cite{Cos} treats in greater detail some of the material in 
the present paper including background on concave functions.

In Section \ref{section:twolevels} of this paper, we state the hypotheses on the statistical
mechanical models to which the theory of the present paper applies,
give a number of examples of such models, and then present the results on ensemble
equivalence at the level of equilibrium macrostates for the three ensembles.  In Section
\ref{section:thermo} we relate ensemble equivalence at the level of equilibrium
macrostates and at the thermodynamic level via the Legendre-Fenchel transform
and a mild generalization suitable for treating quantities arising in the
generalized canonical ensemble.  In Section 4 we present a number of results
giving conditions for the existence of a generalized canonical ensemble that is universally
equivalent to the microcanonical ensemble.  In all but one of these results
the generalized canonical ensemble is Gaussian.  The appendix contains 
a number of technical results on concave functions needed in the
main body of the paper.

\section{Definitions of Models and Ensembles}
\setcounter{equation}{0}
\label{section:definitions}

The main contribution of this paper is that when the canonical ensemble
is nonequivalent to the microcanonical ensemble on a subset of values of $u$, 
it can often be replaced by a generalized canonical ensemble that is equivalent
to the microcanonical ensemble at all $u$.  Before introducing the various
ensembles as well as the methodology for proving this result, we first
specify the class of statistical mechanical models under consideration.
The models are defined in terms of the following quantities.  

\begin{itemize}
\item A sequence of probability spaces $(\Omega_n, {\cal F}_n,P_n)$
indexed by $n \in \N$, which typically represents a sequence of finite dimensional systems.  
The $\Omega_n$ are the configuration spaces, 
$\omega \in \Omega_n$ are the microstates, and the $P_n$ are the prior measures.

\item A sequence of positive scaling constant $a_n \goto \infty$ as $n \goto \infty$.  
In general $a_n$ equals the total number of degrees of freedom in the model.
In many cases $a_n$ equals the number of particles.

\item A positive integer $\sigma$ and for each $n \in \N$ measurable
functions $H_{n,1}, \ldots, H_{n,\sigma}$ mapping $\Omega_n$ into $\R$.  
For $\omega \in \Omega_n$
we define
\[
h_{n,i}(\omega) = \frac{1}{a_n} H_{n,i}(\omega) \ \mbox{ and } 
\ h_n(\omega) = (h_{n,1}(\omega), \ldots, h_{n, \sigma}(\omega)).
\] 
The $H_{n,i}$ include the Hamiltonian and, if $\sigma \geq 2$, other dynamical
invariants associated with the model.
\end{itemize}

A large deviation analysis of the general model is possible provided 
that we can find, as specified in the next four items, a space of macrostates, 
macroscopic variables, and interaction representation functions and provided
that the macroscopic variables satisfy the large deviation principle
(LDP) on the space of macrostates.  

\begin{enumerate}

\item {\bf Space of macrostates}.  This is a complete, separable metric space ${\cal X}$,
which represents the set of all possible macrostates.

\item {\bf Macroscopic variables.}  These are a sequence of random variables
$Y_n$ mapping $\Omega_n$ into ${\cal X}$.  These functions associate a
macrostate in $\X$ with each microstate $\omega \in \Omega_n$.

\item {\bf Interaction representation functions.}  These are
bounded, continuous functions $\tilde{H}_1, \ldots,$ $\tilde{H}_\sigma$ 
mapping ${\cal X}$ into $\R$ such that as $n \rightarrow \infty$
\be 
\label{eqn:interaction}
h_{n,i}(\omega)=\tilde{H}_i(Y_n(\omega)) + \mbox{o}(1) \ \ \mbox{ uniformly for }
\omega \in \Omega_n;
\ee
i.e., 
\[
\lim_{n  \rightarrow \infty} \sup_{\omega \in \Omega_n} 
|h_{n,i}(\omega)-\tilde{H}_i(Y_n(\omega))| = 0.
\]  
We define $\tilde{H}=(\tilde{H}_1, \ldots, \tilde{H}_\sigma)$.
The functions $\tilde{H}_i$ 
enable us to write the $h_{n,i}$, either exactly or asymptotically,
as functions of the macrostate via the macroscopic variables $Y_n$.  

\item {\bf LDP for the macroscopic variables.}  There exists a 
function $I$ mapping ${\cal X}$ into $[0, \infty]$ and having
compact level sets such that with respect to
$P_n$ the sequence $Y_n$ satisfies the LDP on ${\cal X}$ with rate function $I$
and scaling constants $a_n$
In other words, for any closed subset $F$ of $\X$ 
\[
\limsup_{n \goto \infty} \frac{1}{a_n} \log P_n\{Y_n \in F\} \leq - \inf_{x \in F}I(x),
\]
and for any open subset $G$ of $\X$ 
\[
\liminf_{n \goto \infty} \frac{1}{a_n} \log P_n\{Y_n \in G\} \geq - \inf_{x \in G}I(x).
\]
It is helpful to summarize the LDP by the formal notation
$P_n\{Y_n \in dx\} \asymp \exp[-a_n I(x)]$.   This notation expresses the fact
that, to a first degree of approximation, $P_n\{Y_n \in dx\}$ behaves like an
exponential that decays to 0 whenever $I(x) > 0$.
\end{enumerate}

As specified in item 3, the functions $\tilde{H}_i$
are bounded on $\X$, and because of (\ref{eqn:interaction}) the functions
$h_{n,i}$ are also bounded on $\X$.  In \cite{Cos} it is shown that all
the results in this paper are valid under much weaker hypotheses on $\tilde{H}_i$,
including $\thi$ that are not bounded on $\X$.  

The assumptions on the statistical mechanical models just stated 
as well as a number of definitions to follow
are valid for lattice spin and other models.  These assumptions
differ slightly from
those in \cite{EHT1}, where they are adapted for applications to
statistical mechanical models of coherent structures in turbulence.  
The major difference is that $H_n$ in \cite{EHT1}
is replaced by $h_n$ here in several equations: the asymptotic relationship 
(\ref{eqn:interaction}), the definition (\ref{eqn:microens})
of the microcanonical ensemble $P_n^{u,r}$,  
and the definition (\ref{eqn:canonens}) of the canonical ensemble $P_{n,\beta}$.   
In addition, in \cite{EHT1}
the LDP for $Y_n$ is studied with respect to $P_{n,a_n \beta}$, in which 
$\beta$ is scaled by $a_n$; 
here the LDP for $Y_n$ is studied with respect to $P_{n,\beta}$.  
With only such superficial changes
in notation, all the results in \cite{EHT1} are applicable here, and, in turn, 
all the results derived here are applicable to the models considered in \cite{EHT1}.

A wide variety of statistical mechanical models satisfy the hypotheses
listed at the start of this section
and so can be studied by the methods of \cite{EHT1} and the present paper.
We next give six examples.
The first two are long-range spin systems, the third a class of short-range spin
systems, the fourth a model of two-dimensional turbulence, the fifth
a model of quasi-geostrophic turbulence, and the 
sixth a model of dispersive wave turbulence.

\bexa
\label{exa:six}

\noi
{\bf 1. Mean-field Blume-Emery-Griffiths model.}  The Blume-Emery
Griffiths model \cite{BEG} is one of the few and certainly one of the simplest lattice-spin
models known to exhibit, in the mean-field approximation, both a continuous, second-order phase transition
and a discontinuous, first-order phase transition.   This mean-field model 
is defined on the set $\{1,2,\ldots,n\}$.  The spin at site $j \in \{1,2,\ldots,n\}$
is denoted by $\omega_j$, a quantity taking values in $\Lambda = \{-1,0,1\}$.  The configuration
spaces for the model are $\Omega_n = \Lambda^n$, the prior measures $P_n$
are product measures on $\Omega_n$ with identical one-dimensional marginals
$\rho = \frac{1}{3}(\delta_{-1} + \delta_0 + \delta_{1})$, and 
for $\omega = (\omega_1,\ldots,\omega_n) \in \Omega_n$ the Hamiltonian
is given by
\[
H_n(\omega) = \sum_{j=1}^n \omega_j^2 - \frac{K}{n} \!\left(\sum_{j=1}^n \omega_j\right)^2,
\]
where $K$ is a fixed positive number.  
The space of macrostates for this model is the set of probability measures on $\Lambda$,
the macroscopic variables are the empirical measures associated with the spin configurations
$\omega$, and the associated 
LDP is Sanov's Theorem, for which the rate function is the relative entropy with respect to $\rho$.
The large deviation analysis of the model is given
in \cite{EllOttTou}, which also analyzes the phase transition in the model.  
Equivalence and nonequivalence of ensembles for this model is studied 
at the thermodynamic level in \cite{BMR1,BMR2,ETT} and at the level of equilibrium macrostates
in \cite{ETT}.

\skp
\noi
{\bf 2. Curie-Weiss-Potts model.}  The Curie-Weiss-Potts model is a long-range,
mean-field approximation to the well known Potts model \cite{Wu}.
It is defined on the set $\{1,2,\ldots,n\}$.  The spin at site $j \in \{1,2,\ldots,n\}$
is denoted by $\omega_j$, a quantity taking values in the set $\Lambda$ consisting
of $q$ distinct vectors $\theta^i \in \r^q$, where $q \geq 3$ is a fixed integer.  
The configuration spaces for the model are $\Omega_n = \Lambda^n$, the prior measures $P_n$
are product measures on $\Omega_n$ with identical one-dimensional marginals
$\frac{1}{q}\sum_{i=1}^q \delta_{\theta^i}$, and 
for $\omega = (\omega_1,\ldots,\omega_n) \in \Omega_n$ the Hamiltonian
is given by
\[
H_n(\omega) = -\frac{1}{2n}\sum_{j,k=1}^n \delta(\omega_j,\omega_k).
\]
As in the case of the mean-field Blume-Emery-Griffiths model,
the space of macrostates for the Curie-Weiss-Potts model is the set of probability
measures on $\Lambda$, the macroscopic variables are the empirical measures associated with $\omega$,
and the associated LDP is Sanov's Theorem, for which the rate function
is the relative entropy with respect to $\rho$.
The large deviation analysis of the model
is summarized in \cite{CosEllTou1}, which together with \cite{CosEllTou2} gives a complete analysis
of ensemble equivalence and nonequivalence at the level of equilibrium macrostates.

\skp
\noi
{\bf 3. Short-range spin systems.}  
Short-range spin systems such as the Ising model on $\ZZ^d$
and numerous generalizations can also be handled by the methods of this paper.  The
large deviation techniques required to analyze these models
are much more subtle than in the case of 
the long-range, mean-field models considered in items 1 and 2.
The already complicated large deviation analysis of one-dimensional models is given in Section IV.7 of
\cite{Ell}.  The even more sophisticated analysis of multi-dimensional models is
carried out in \cite{FoeOre,Olla}.   For these spin systems the space of macrostates 
is the space of translation-invariant probability measures on $\ZZ^d$, the macroscopic
variables are the empirical processes associated with the spin configurations, 
and the rate function in the associated LDP the
mean relative entropy.

\skp
\noi
{\bf 4. A model of two-dimensional turbulence.}
The Miller-Robert model is a model of coherent structures in an ideal,
two-dimensional fluid that includes all the exact invariants of the
vorticity transport equation \cite{Miller,Robert}.  In its original formulation,
the infinite family of enstrophy integrals is imposed
microcanonically along with the energy.  If this formulation is
slightly relaxed to include only finitely many enstrophy integrals,
then the model can be put in the general form described above; that
form can also be naturally extended to encompass complete enstrophy
conservation.  The space of macrostates is the space of Young measures
on the vorticity field; that is, a macrostate has the form
$\nu(x,dz)$, where $x \in \Lambda$ runs over the fluid domain
$\Lambda$, $z$ runs over the range of the vorticity field
$\zeta(x)$, and for almost all $x$, $n(x,dz)$ is a probability measure in
$z$.  The large deviation analysis of this model developed first in \cite{Robert}
and more recently in \cite{BouEllTur} gives a rigorous derivation of maximum entropy principles
governing the equilibrium behavior of the ideal fluid.

\skp
\noi
{\bf 5. A model of quasi-geostrophic turbulence.}
In later formulations, especially in geophysical applications, another
version of the model in item 4 is preferred, in which the enstrophy integrals
are treated canonically and the energy and circulation are treated
microcanonically \cite{EHT2}.  In those formulations, the space of
macrostates is $L^2(\Lambda)$ or $L^{\infty}(\Lambda)$ depending on
the contraints on the voriticty field.  The large deviation analysis
for such a formulation is carried out in \cite{EHT3}.  Numerical results
given in \cite{EHT2} illustrate key examples of nonequivalence with respect
to the energy and circulation invariants.  In addition, this paper
shows how the nonlinear stability of the steady mean flows arising as
equilibriums macrostates in these models can be established by
utilizing the appropriate generalized thermodynamic potentials.

\skp
\noi
{\bf 6.  A model of dispersive wave turbulence.}  A statistical
equilibrium model of solitary wave structures in dispersive wave
turbulence governed by a nonlinear Schr\"odinger equation is studied in
\cite{EllJorOttTur}.  In this model the energy is treated canonically while the
particle number invariant is imposed microcanonically; without the
microcanonical constraint on particle number the ensemble is not
normalizable for focusing nonlinearities.  The large deviation analysis
given in \cite{EllJorOttTur} 
derives rigorously the concentration phenomenon observed in long-time
numerical simulations and predicted by mean-field approximations
\cite{JorTurZir,LebRosSpe2}.  The space of macrostates
is $L^2(\Lambda)$, where $\Lambda$ is a bounded interval or more
generally a bounded domain in $\R^d$.  \ \ink
\eexa

We now return to the general theory, first
introducing the function whose support and concavity properties 
completely determine all aspects of ensemble equivalence and nonequivalence.
This function is the microcanonical entropy, defined for $u \in \rsigma$ by
\be 
\label{eqn:entropy}
s(u) = -\inf \{I(x) : x \in {\cal X}, \tilde{H}(x) = u\}.
\ee
Since $I$ maps ${\cal X}$ into $[0,\infty]$, $s$ maps
$\rsigma$ into $[-\infty,0]$.  Moreover, since $I$ is lower semicontinuous
and $\thi$ is continuous on $\X$, $s$ is 
upper semicontinuous on $\rsigma$.
We define $\mbox{dom} \, s$ to be the set of $u \in \rsigma$ for which $s(u) > -\infty$.
In general, $\mboxdoms$ is nonempty since $-s$ is a rate function
\cite[Prop.\ 3.1(a)]{EHT1}.  
For each $u \in \mbox{dom} \, s$, $r > 0$,
$n \in \N$, and set $B \in {\cal F}_n$ the microcanonical ensemble 
is defined to be the conditioned measure
\be
\label{eqn:microens}
P^{u,r}_n \{B\}=P_n \{B \mid h_n \in \{u\}^{(r)}\},
\ee
where $\{u\}^{(r)}=[u_1-r,
u_1+r]\times \cdots \times[u_\sigma -r, u_\sigma +r]$.
As shown in \cite[p.\ 1027]{EHT1},
if $u \in \mbox{dom} \, s$, then for all sufficiently large $n$,
$\, P_n\{h_n \in \{u\}^{(r)}\} > 0$; thus the
conditioned measures  $P^{u,r}_n $ are well defined.

A mathematically more tractable probability measure is
the canonical ensemble.               
Let $\langle \cdot, \cdot \rangle$ denote the Euclidian inner
product on $\R^\sigma$.  For each $n \in \N$, $\beta
\in \R^\sigma$, and set $B \in {\cal F}_n$ we define the partition function
\[ 
Z_n(\beta) = \int_{\Omega_n}  \exp [-a_n\langle \beta, h_n \rangle] \, dP_n,
\]
which is well defined and finite, and the probability measure
\be
\label{eqn:canonens}
P_{n, \beta}\{B\} = \frac{1}{Z_n(\beta)} \cdot \int_{B} \exp [-a_n \langle
\beta, h_n \rangle] \, dP_n.
\ee
The measures $P_{n, \beta}$ are Gibbs states that define the canonical
ensemble for the given model.  

The generalized canonical ensemble is a natural perturbation of the
canonical ensemble, defined in terms of a continuous function
$g$ mapping $\R^\sigma$ into $\R$.  For each $n \in \N$
and $\beta \in \rsigma$ we define
the generalized partition function
\be
Z_{n,g}(\beta)=\int_{\Omega_n}  \exp [-a_n\langle \beta, h_n
\rangle -a_n g(h_n)] \, dP_n.
\ee
This is well defined and finite because the $h_n$ are bounded
and $g$ is bounded on the range of the $h_n$.  For $B \in {\cal F}_n$ we
also define the probability measure
\be
\label{eqn:gencanon} 
P_{n, \beta,g}\{B\} = \frac{1}{Z_{n,g}(\beta)} \cdot \int_{B}
\exp [-a_n\langle \beta, h_n \rangle - a_n g(h_n)] \, dP_n,
\ee
which we call the generalized canonical ensemble.
The special case in which $g$ equals a quadratic function
gives rise to the Gaussian ensemble \cite{CH1,CH2,Heth1987,Stump1987,JPV,Stump21987}.

In order to define the sets of equilibrium macrostates
for each ensemble, we summarize two large deviation
results proved in \cite{EHT1} and extend one of them.  
It is proved in \cite[Thm.\ 3.2]{EHT1} 
that with respect to the microcanonical ensemble
$P_n^{u,r},$ $Y_n$ satisfies the LDP on ${\cal X}$, in the
double limit $n \rightarrow \infty$ and $r \rightarrow 0$, with rate
function
\be
\label{eqn:iu}
I^u(x) = \left\{
\begin{array}{ll}
           I(x)+s(u) & \ \mbox{ if } \: \tilde{H}(x)=u
\\
           \infty & \ \mbox{ otherwise. }
\end{array}
\right.
\ee
$I^u$ is nonnegative on $\X$, and for $u \in \mboxdoms$, $I^u$ 
attains its infimum of 0 on the set
\bea
\label{eqn:eu}
 {{\cal E}}^u & = & \{x \in {\cal X} : I^u(x)=0\} \\
& = & \{x \in \X : I(x) \mbox{ is minimized subject to } \thi(x) = u\}.
\nonumber 
\eea

In order to state the LDPs for the other two ensembles, we 
bring in the canonical free energy, defined for $\beta \in \rsigma$ by 
\[
\varphi(\beta) = - \lim_{n \goto \infty} \frac{1}{a_n}
\log Z_{n}(\beta),
\]
and the generalized canonical free energy, defined by 
\[
\varphi_g(\beta) = - \lim_{n \goto \infty} \frac{1}{a_n}
\log Z_{n,g}(\beta).
\]
Clearly $\varphi_0(\beta) = \varphi(\beta)$. 
It is proved in \cite[Thm.\ 2.4]{EHT1} that the limit defining $\varphi(\beta)$ exists
and is given by
\be
\label{eqn:varphi}
\varphi(\beta) =  \inf_{y \in {\cal X}} \{I(y) + \langle \beta, \tilde{H}(y)\ran\}
\ee
and that with respect to $P_{n,\beta}$, $Y_n$ satisfies
the LDP on ${\cal X}$ with rate function
\be
\label{eqn:ibeta}
I_\beta(x)=I(x)+ \langle \beta, \tilde{H}(x) \rangle - \varphi(\beta).
\ee
$I_\beta$ is nonnegative on $\X$ and attains its infimum of 0 on
the set
\bea
\label{eqn:ebeta}
{{\cal E}}_\beta & =& \{x \in {\cal X} : I_\beta(x)=0\} \\
& = & \{x \in \X : I(x) + \lan \beta,\thi(x) \ran \mbox{ is minimized}\}. \nonumber
\eea

A straightforward extension of these results shows
that the limit defining $\varphi_g(\beta)$ exists and is given by
\be
\label{eqn:varphig}
\varphi_g(\beta) = \inf_{y \in \X}\{I(y) + \lan \beta,\thi(y)\ran  + g(\thi(y))\}
\ee
and that with respect to $P_{n,\beta,g}$, $Y_n$ satisfies
the LDP on ${\cal X}$ with rate function
\be
\label{eqn:ibetag}
I_{\beta,g}(x)=I(x)+ \langle \beta, \tilde{H}(x) \rangle + g(\thi(x)) - \varphi_g(\beta).
\ee
$I_{\beta,g}$ is nonnegative on $\X$ and attains its infimum of 0 on
the set
\bea
\label{eqn:egbeta}
{{\cal E}}(g)_\beta & = & \{x \in {\cal X} : I_{\beta,g}(x)=0\} \\
& = &  \{x \in \X : I(x) + \lan \beta,\thi(x) \ran + g(\thi(x)) 
\mbox{ is minimized}\}. \nonumber
\eea

For $u \in \mboxdoms$, let $x$ be any element of $\X$ satisfying
$I^u(x) > 0$.  The formal notation
\[
P_{n}^{u,r}\{Y_n \in dx\} \asymp e^{-a_n I^u(x)}
\]
suggests that $x$ has an exponentially small probability of
being observed in the limit  $n \goto \infty$, $r \goto 0$.  
Hence it makes sense to identify $\eu$
with the set of microcanonical equilibrium macrostates.  In the same way we identify
with $\ebeta$ the set of canonical equilibrium macrostates and with 
$\egbeta$ the set of generalized canonical equilibrium macrostates.
A rigorous justification is given in \cite[Thm.\ 2.4(d)]{EHT1}.

\section{Ensemble Equivalence at the Level of Equilibrium Macrostates}
\setcounter{equation}{0}
\label{section:twolevels}

Having defined the sets
of equilibrium macrostates $\eu$, $\ebeta$, and $\egbeta$ for the microcanonical,
canonical and generalized canonical ensembles, we now come to the main
point of this paper, which is to show how these sets relate to one
another.  In Theorem \ref{thm:equiv} we state the results proved
in \cite{EHT1} concerning equivalence and nonequivalence at the level
of equilibrium macrostates for the microcanonical
and canonical ensembles.  Then in Theorem \ref{thm:main}
we extend these results to the generalized canonical ensemble.  

Parts (a)--(c) of Theorem \ref{thm:equiv}
give necessary and sufficient conditions, in terms of 
support properties of $s$, 
for ensemble equivalence and nonequivalence of $\eu$ and $\ebeta$.
These assertions are proved in Theorems 4.4 and 4.8 in \cite{EHT1}.
Part (a) states that $s$ has a 
strictly supporting hyperplane at $u$ if and only if full equivalence of ensembles
holds; i.e., if and only if there exists a $\beta$ such that
$\eu = \ebeta$.  The most surprising result, given in part (c), is 
that $s$ has no supporting hyperplane at $u$ if and only if nonequivalence of ensembles
holds in the strong sense that $\eu \cap \ebeta = \emptyset$ for 
all $\beta \in \rsigma$.  
Part (c) is to be contrasted with part (d), which
states that for any $\beta \in \rsigma$ 
canonical equilibrium macrostates can always be realized
microcanonically.  Part (d) is proved in Theorem 4.6 in \cite{EHT1}.
Thus one conclusion of this theorem is that at the level
of equilibrium macrostates the microcanonical
ensemble is the richer of the two ensembles.  
The concept of a relative boundary point, which arises in part (c),
is defined after the statement of the theorem.
For $\beta \in \rsigma$, $[\beta,-1]$ denotes the vector 
in $\R^{\sigma + 1}$ whose first $\sigma$ components agree
with those of $\beta$ and whose last component equals $-1$.

\begin{theorem}\per  In parts {\em (a)}, {\em (b)}, and {\em (c)},
$u$ denotes any point in $\mbox{{\em dom}} \, s$.
\label{thm:equiv}  

{\em (a)}  \mbox{{\em {\bf Full equivalence.}}}  \ There exists $\beta \in \rsigma$
such that ${\cal E}^u = {\cal E}_\beta$ 
if and only if $s$ has a strictly supporting hyperplane at $u$ with normal vector 
$[\beta,-1]$; i.e.,
\[
s(v) < s(u) + \langle \beta, v-u \rangle \mbox{ for all } v \not = u.
\]

{\em (b)} \mbox{{\em {\bf Partial equivalence.}}} \ There exists $\beta \in \rsigma$
such that
${\cal E}^u \subset {\cal E}_\beta$ but ${\cal E}^u \not = {\cal E}_\beta$
if and only if $s$ has a nonstrictly supporting hyperplane at $u$ with normal vector 
$[\beta,-1]$; i.e., 
\[
s(v) \leq  s(u) + \langle \beta, v-u \rangle \mbox{ for all } v 
\mbox{ with equality for some } v \not = u.
\]

{\em (c)} \mbox{{\em {\bf Nonequivalence.}}} \ For all $\beta \in \rsigma$,
${\cal E}^u \cap {\cal E}_\beta = \emptyset$
if and only if $s$ has no supporting hyperplane at $u$; i.e., 
\[
\mbox{for all } \beta \in \rsigma \mbox{there exists } v \mbox{ such that }
s(v) > s(u) + \lan \beta,v-u \ran.
\]  
Except possibly for relative boundary points of $\mboxemdoms$,
the latter condition is equivalent to the nonconcavity of $s$ at $u$
{\em [Thm.\ \ref{thm:5.7}(c)]}. 

{\em (d)}  \mbox{{\em {\bf Canonical is always realized microcanonically.}}}
\ For any $\beta \in \rsigma$ we have $\tilde{H}({\cal E}_\beta) \subset 
\mbox{\em dom} \, s$ and
\[
{\cal E}_\beta = \bigcup_{u \in \tilde{H}({\cal E}_\beta)} {\cal E}^u.
\]
\end{theorem}

We highlight several features of the theorem in order to illuminate
their physical content.  In part (a) we assume that for a given $u \in \mboxdoms$
there exists a unique $\beta$ such that $\eu = \ebeta$.  
If $s$ is differentiable at $u$ and $s$ and the double-Legendre-Fenchel transform $s^{**}$
are equal in a neighborhood of $u$, then $\beta$ is given by the
standard thermodynamic formula $\beta = \nabla s(u)$ [Thm.\ \ref{prop:2.9}(b)].  The inverse
relationship can be obtained from part (d) of the theorem under
the assumption that $\ebeta$ consists of a unique macrostate
or more generally that for all $x \in \ebeta$ the values
$\tilde{H}(x)$ are equal.  Then $\ebeta = {\cal E}^{u(\beta)}$, where
$u(\beta) = \tilde{H}(x)$ for any $x \in \ebeta$; $u(\beta)$
denotes the mean energy realized at equilibrium in the canonical ensemble.  The relationship
$u = u(\beta)$ inverts the relationship $\beta = \nabla s(u)$.  Partial ensemble
equivalence can be seen in part (d) under the assumption that for a given $\beta$,
$\ebeta$ can be partitioned into at least two sets ${\cal E}_{\beta,i}$ such that
for all $x \in {\cal E}_{\beta,i}$ the values $\tilde{H}(x)$ are equal but 
$\tilde{H}(x) \not = \tilde{H}(y)$
whenever $x \in {\cal E}_{\beta,i}$ and $y \in {\cal E}_{\beta,j}$ for $i \not = j$. Then
$\ebeta = \bigcup_{i}{\cal E}^{u_i(\beta)}$, where $u_i(\beta) = \tilde{H}(x)$, $x \in {\cal E}_{\beta,i}$.
Clearly, for each $i$, ${\cal E}^{u_i(\beta)} \subset \ebeta$ but ${\cal E}^{u_i(\beta)} 
\not = \ebeta$.  Physically, this
corresponds to a situation of coexisting phases that normally
takes place at a first-order phase transition \cite{Touchette2004}. 

Theorem 4.10 in \cite{EHT1} states an alternative version of part (d) of Theorem \ref{thm:equiv}, 
in which the set $\tilde{H}({\cal E}_\beta)$
of canonical equilibrium mean-energy values is replaced by another set.
We next present a third version of part (d) that could be useful in applications.
This corollary is also aesthetically pleasing because like parts (a)--(c)
of Theorem \ref{thm:equiv} it is formulated in terms of 
support properties of $s$.

\begin{corollary}\per
\label{cor:third}
For $\beta \in \rsigma$ we define
$A_\beta$ to be the set of $u \in \mbox{{\em dom}} \, s$ such that $s$ 
has a supporting hyperplane at $u$ with 
normal vector $[\beta, -1]$.  Then
\[
{\cal{E}}_\beta = \bigcup_{u \in A_\beta} {\cal{E}}^u.
\]
\end{corollary}

\noi
{\bf Proof.} Part (d) of Theorem \ref{thm:equiv} implies 
that if $u \in \tilde H({\cal{E}}_\beta)$, then
${\cal{E}}^u \subset {\cal{E}}_\beta$.  From parts (a) and (b) of the theorem 
it follows that $s$ has a supporting hyperplane at $u$ with normal vector $[\beta, -1]$.  
Hence $\thi(\ebeta) \subset A_\beta$ and 
\[
{\cal{E}}_\beta = \bigcup_{u \in \thi(\ebeta)} {\cal{E}}^u 
\subset \bigcup_{u \in A_\beta} {\cal{E}}^u.
\]
The reverse inclusion is also a consequence of 
parts (a) and (b) of the theorem, which imply
that if $u \in A_\beta$, then ${\cal{E}}^u \subset {\cal{E}}_\beta$
and thus that
\[
\bigcup_{u \in A_\beta} {\cal{E}}^u \subset {\cal{E}}_\beta.
\]
This completes the proof. \ \ink

\skp

Before continuing with our analysis of ensemble equivalence, we introduce several sets
that play a central role in the theory.
Let $f \not \equiv -\infty$ be a function mapping $\rsigma$ into $\r \cup \{-\infty\}$.
The relative interior of $\mboxdomf$,
denoted by $\mboxridomf$, is defined as the interior of $\mboxdomf$ when considered
as a subset of the smallest affine set that contains $\mboxdomf$.  
Clearly, if the smallest affine
set that contains $\mboxdomf$ is $\rsigma$, then the relative
interior of $\mboxdomf$ equals the interior of $\mboxdomf$,
which we denote by $\mbox{int(dom} \, f)$. 
This is the case if, for example, $\sigma = 1$ and $\mboxdomf$
is a nonempty interval.
The relative boundary of $\mboxdomf$ is defined as
$\mbox{cl(dom} \, f) \setminus \mbox{ri(dom} \, f)$.   

We continue by giving several definitions for concave functions on $\rsigma$
when $\sigma$ is an arbitrary positive integer.  We then
 specialize to the case $\sigma = 1$, for which all the concepts
can be easily visualized.  Additional material on concave
functions is contained in the appendix.  Let $f$ be a concave function on $\rsigma$.  
For $u \in \rsigma$ the superdifferential of $f$ at $u$, denoted by $\partial f(u)$,
is defined to be the set of $\beta \in \rsigma$ such that $[\beta,-1]$ is the normal
vector to a supporting hyperplane of $f$ at $u$; i.e.,
\[
f(v) \leq f(u) + \lan \beta,v-u \ran \mbox{ for all } v \in \rsigma.
\] 
Any such $\beta$ is called a supergradient of $f$ at $u$.  The domain of $\partial f$,
denoted by $\mbox{dom}\, \partial f$, is then defined to be the set of $u$ 
for which $\partial f(u) \not = \emptyset$.  A basic fact is that $\mbox{dom}\, \partial f$ is
a subset of $\mboxdomf$ and differs from it, if at all, only in a subset of the relative boundary of 
$\mboxdomf$; a precise statement is given in part (a) of Theorem \ref{thm:usefulfacts}.
By definition of $\mbox{dom} \, \partial f$, it follows
that $f$ has a supporting hyperplane at all points of $\mboxdomf$ except
possibly relative boundary points. 

We now specialize to the case $\sigma = 1$, considering a concave function
$f$ mapping $\r$ into $\r \cup \{-\infty\}$ for which $\mbox{dom} \, f$ 
is a nonempty interval $L$.
For $u \in L$, $\partial f(u)$ is defined to be the set 
of $\beta \in \r$ such that $\beta$ is the slope of a supporting
line of $f$ at $u$.  Thus, if $f$ is differentiable at $u \in \mbox{int} \, L$,
then $\partial f(u)$ consists of the unique point $\beta = f'(u)$.
If $f$ is not differentiable at $u \in \mbox{int} \, L$, 
then $\mbox{dom}\, \partial f$
consists of all $\beta$ satisfying the inequalities 
\[
(f')^+(u) \leq \beta \leq (f')^-(u),
\]
where $(f')^-(u)$ and $(f')^+(u)$ denote the left-hand and right-hand
derivatives of $f$ at $u$.  

Complications arise because 
$\mbox{dom} \, \partial f$ can be a proper subset of $\mboxdomf$,
as the situation in one dimension clearly shows.  Let $b$ be a
boundary point of $\mboxdomf$ for which $f(b) > -\infty$.  Then $b$ is in
$\mbox{dom} \, \partial f$ if and only if the one-sided derivative of $f$ at $b$ is
finite.  For example, if $b$ is a left hand boundary point of $\mboxdomf$ and $(f')^+(b)$ is finite,
then $\partial f(b) = [(f')^+(b),\infty)$; any $\beta \in \partial f(b)$ is
the slope of a supporting line at $b$.
The possible discrepancy
between $\mbox{dom} \, \partial f$ and $\mboxdomf$ introduces unavoidable technicalities 
in the statements of many
results concerning the existence of supporting hyperplanes.  

One of our goals is to find concavity and support
conditions on the microcanonical entropy guaranteeing that
the microcanonical and canonical 
ensembles are fully equivalent at all points $u \in \mbox{dom} \, s$ except possibly
relative boundary points.  If this is the case, then we say that the ensembles are
{\bf universally equivalent}.  Here is a basic result in that direction. 

\begin{thm}\per
\label{thm:strictlyconcave}
Assume that $\mboxemdoms$ is a convex subset of $\rsigma$ and that
$s$ is strictly concave on $\mboxemridoms$ and continuous on $\mboxdoms$.
The following conclusions hold.

{\em (a)} $s$ has a strictly supporting hyperplane at all
$u \in \mboxemdoms$ except possibly relative boundary points.

{\em (b)} The microcanonical and canonical ensembles are universally 
equivalent; i.e., fully equivalent at all $u \in \mboxemdoms$ 
except possibly relative boundary points.

{\em (c)} $s$ is concave on $\rsigma$, and 
for each $u$ in part {\em (b)} the corresponding $\beta$ in the statement of
full equivalence is any element of $\partial s(u)$.  

{\em (d)} If $s$ is differentiable
at some $u \in \mboxemdoms$, then the corresponding $\beta$ in part {\em (b)} is unique and is given by
the standard thermodynamic formula $\beta = \nabla s(u)$.
\end{thm}

\noi
{\bf Proof.}  (a) This is a consequence of part (c) of Theorem \ref{prop:2.9}.  

(b) The universal equivalence follows from part (a) of Theorem \ref{thm:equiv}.

(c)  By Proposition \ref{prop:extend}
the continuity of $s$ on $\mboxdoms$ allows us to extend the strict concavity of 
$s$ on $\mboxridoms$ to the concavity of $s$ on $\mboxdoms$.
Since $s$ equals $-\infty$ on the complement 
of $\mboxdoms$, $s$ is also concave on $\rsigma$.  
The second assertion in part (c) is the definition of supergradient.  

(d) This is a consequence of part (c) of the present theorem and 
part (b) of Theorem \ref{thm:usefulfacts}. \ \ink

\skp

We now come to the main result of this paper, which extends Theorem
\ref{thm:equiv} by giving equivalence and
nonequivalence results involving $\eu$ and $\egbeta$.
The proof of the theorem makes it transparent
why $s$ in Theorem \ref{thm:equiv} is replaced here by $s-g$.
In \cite{Cos} an independent proof of Theorem \ref{thm:main} 
is derived from first principles rather than from Theorem \ref{thm:equiv}.  
As we point out after the statement of Theorem \ref{thm:main}, for the purpose
of applications part (a) is its most important contribution.  In order 
to illuminate its physical content, we note that if 
$s-g$ is differentiable at some $u \in \mboxdoms$ and $s-g = (s-g)^{**}$
in a neighborhood of $u$, then $\beta$ is unique and is given by the 
thermodynamic formula $\beta = \nabla (s-g)(u)$ [Thm.\ \ref{prop:2.9}(b)].  

\begin{theorem}\per 
\label{thm:main} 
Let $g$ be a continuous function mapping $\rsigma$ into $\r$, in 
terms of which the generalized canonical ensemble
{\em (\ref{eqn:gencanon})} is defined.
The following conclusions hold.  
In parts {\em (a)}, {\em (b)}, and {\em (c)}, $u$ denotes
any point in $\mbox{\em{dom}} \, s$.

{\em (a)}  \mbox{{\em {\bf Full equivalence.}}} \ There exists $\beta \in \rsigma$
such that ${\cal E}^u = {\cal E}(g)_\beta$ 
if and only if $s - g$ has a strictly supporting hyperplane at $u$ with normal vector
$[\beta,-1]$. 

{\em (b)} \mbox{{\em {\bf Partial equivalence.}}} \ There exists $\beta \in \rsigma$ such that
${\cal E}^u \subset {\cal E}(g)_\beta$ but ${\cal E}^u \not = {\cal E}_\beta$
if and only if $s - g$ has a nonstrictly supporting hyperplane at $u$ with normal vector 
$[\beta,-1]$. 

{\em (c)} \mbox{{\em {\bf Nonequivalence.}}} \ For all $\beta \in \rsigma$,
${\cal E}^u \cap {\cal E}(g)_\beta = \emptyset$
if and only if $s - g$ has no supporting hyperplane at $u$.
Except possibly for relative boundary points of $\mboxemdoms$,
the latter condition is equivalent to the nonconcavity of $s-g$ at $u$
{\em [Thm.\ \ref{thm:5.7}(c)]}.  

{\em (d)} \mbox{{\em {\bf Generalized canonical is always realized microcanonically.}}}
\ For any  $\beta \in R^\sigma$ we have $\tilde{H}({{\cal E}}(g)_\beta)
\subset \mbox{{\em dom}} \, s$ and 
\[
{\cal E}(g)_\beta= \bigcup_{u \in \thi({{\cal E}}(g)_\beta)} 
{\cal E}^u.
\]
\end{theorem}

\noi
{\bf Proof.}  \ For $B \in {\cal F}_n$ we define a new probability measure
\[
P_{n,g}\{B\} = \frac{1}{\displaystyle
\int_{\Omega_n} \exp[-a_n g(h_n)] \, dP_n} \cdot \int_B \exp[-a_n g(h_n)] \, dP_n.
\]
Replacing the prior measure $P_n$
in the standard canonical ensemble with $P_{n,g}$ gives the generalized
canonical ensemble $P_{n,\beta,g}$; i.e., 
\[
P_{n,\beta,g}\{B\} = \frac{1}{\displaystyle
\int_{\Omega_n} \exp[-a_n \lan\beta,h_n\ran] \, dP_{n,g}} 
\cdot \int_B \exp[-a_n \lan\beta,h_n\ran] \, dP_{n,g}.
\]
We also introduce a new conditioned measure
\[
P_{n,g}^{u,r}\{B\} = P_{n,g}\{B \mid h_n \in \{u\}^{(r)}\},
\]
obtained from the microcanonical ensemble $P_{n}^{u,r}$ by
replacing $P_n$ with $P_{n,g}$.  
Since $g$ is continuous, for $\omega$ in the set $\{h_n \in \{u\}^{(r)}\}$,
$g(h_n(\omega))$ converges to $g(u)$ uniformly in $\omega$ and $n$ as $r \goto 0$.  
It follows that with respect to $P_{n,g}^{u,r}$, $Y_n$ satisfies the LDP on $\X$, in 
the double limit $n \goto \infty$ and $r \goto 0$, with the same rate function
$I^u$ as in the LDP for $Y_n$ with respect to $P_n^{u,r}$.  As a result,
the set ${\cal E}(g)^u$ of equilibrium macrostates corresponding to 
$P_{n,g}^{u,r}$ coincides with the set $\eu$ of microcanonical equilibrium macrostates.  

At this point we recall that 
according to Theorem \ref{thm:equiv}, all equivalence and nonequivalence relationships
between $\eu$ and $\ebeta$ are expressed
in terms of support properties of 
\[
s(u) = -\inf\{I(x) : x \in \X, \thi(x) = u\},
\]
where $I$ is the rate function in the LDP for $Y_n$ with respect
to the prior measures $P_n$.  
With respect to the new prior measures $P_{n,g}$, $Y_n$ 
satisfies the LDP on $\X$ with rate function
\[
I_g(x) = I(x) + g(\thi(x)) - \mbox{const}.
\]
It follows that all equivalence and nonequivalence relationships
between ${\cal E}(g)^u$ and $\egbeta$ are expressed
in terms of support properties of the function $s_g$ obtained
from $s$ by replacing the rate function $I$ 
by the new rate function $I_g$.  The function $s_g$ is given by
\beas
s_g(u) & = & - \inf\{I_g(x) : x \in \X, \tilde{H}(x) = u\} \\
& = & -\inf\{I(x) + g(\tilde{H}(x)) : x \in \X, \tilde{H}(x) = u\} + \mbox{const} \\
& = & s(u) - g(u) + \mbox{const}.
\eeas
Since ${\cal E}(g)^u = \eu$ and since
$s_g$ differs from $s-g$ by a constant, we conclude that all equivalence and nonequivalence
relationships between $\mathcal{E}^u$ and $\mathcal{E}(g)_\beta $ are expressed
in terms of the same support properties of $s-g$. 
This completes the derivation of
Theorem \ref{thm:main} from Theorem \ref{thm:equiv}. 
\ \ink

\skp
The relationships between
${{\cal E}}^u$ and ${{\cal E}}(g)_\beta$ in Theorem \ref{thm:main}
are valid under much weaker assumptions on both $g$ and $\tilde{H}_i$
that guarantee that these sets are nonempty.  For example, the continuity
of $g$ is not needed.
Of course, if one does not have the LDPs for $Y_n$ with respect to
$P_{n}^{u,r}$ and $P_{n, \beta,g}$, then one cannot
interpret ${{\cal E}}^u$ and ${{\cal E}}(g)_\beta$ as sets of
equilibrium macrostates for the two ensembles.  A similar comment applies
to Theorem \ref{thm:equiv}.

The next corollary gives an alternative version of part (d) of Theorem \ref{thm:main}.
It follows from the theorem in the same way that Corollary \ref{cor:third} follows
from Theorem \ref{thm:equiv}, which is the analogue of Theorem \ref{thm:main}
for the canonical ensemble.

\begin{corollary}\per
\label{cor:thirdg}
Let $g$ be a continuous function mapping $\rsigma$ into $\r$, in 
terms of which the generalized canonical ensemble 
{\em (\ref{eqn:gencanon})} is defined.
For $\beta \in \rsigma$ we define
$A(g)_\beta$ to be the set of $u \in \mbox{{\em dom}} \, s$ such that $s-g$ 
has a supporting hyperplane at $u$ with 
normal vector $[\beta, -1]$.  Then
\[
{\cal{E}}(g)_\beta = \bigcup_{u \in A(g)_\beta} {\cal{E}}^u.
\]
\end{corollary}

The importance of part (a) of Theorem \ref{thm:main} in applications
is emphasized by the following theorem, which will be applied
several times in the sequel.  
This theorem is the analogue of Theorem \ref{thm:strictlyconcave} for
the generalized canonical ensemble, replacing $s$ in that theorem with $s-g$.
Since $g$ takes values in $\R$, the domain of $s-g$ equals the domain of $s$.
Theorem \ref{thm:sgstrictlyconcave} is proved exactly like Theorem \ref{thm:strictlyconcave}. 

\begin{thm}\per
\label{thm:sgstrictlyconcave}
Assume that $\mboxemdoms$ is a convex subset of $\rsigma$ and that
$s-g$ is strictly concave on $\mboxemridoms$ and continuous on $\mboxdoms$.
The following conclusions hold.

{\em (a)} $s-g$ has a strictly supporting hyperplane at all
$u \in \mboxemdoms$ except possibly relative boundary points.

{\em (b)} The microcanonical and generalized canonical ensembles are universally 
equivalent; i.e., fully equivalent at all $u \in \mboxemdoms$ 
except possibly relative boundary points.  

{\em (c)} $s-g$ is concave on $\rsigma$, and
for each $u$ in part {\em (b)} the corresponding $\beta$ in the statement of
full equivalence is any element of $\partial (s-g)(u)$.  

{\em (d)} If $s-g$ is differentiable
at some $u \in \mboxemdoms$, then the corresponding $\beta$ in 
part {\em (b)} is unique and is given by
the thermodynamic formula $\beta = \nabla (s-g)(u)$.
\end{thm}

The most important repercussion of Theorem \ref{thm:sgstrictlyconcave} is the ease
with which one can prove that the microcanonical and generalized 
canonical ensembles are universally equivalent in those cases 
in which microcanonical and standard canonical
ensembles are not fully or partially equivalent.  In order to achieve universal equivalence, 
one merely chooses $g$ so that $s-g$ is strictly concave on 
$\mboxridoms$.  One has considerable
freedom doing this since the only requirement is that $g$ be continuous.
Section \ref{section:gencanon} is devoted to this and related issues.
In Theorems \ref{thm:applysigma1}--\ref{thm:localapply}
we will give several useful examples, three of which involve quadratic functions $g$.

In the next section we introduce the thermodynamic level of ensemble
equivalence and discuss its relationship to ensemble equivalence at the 
level of equilibrium macrostates.

\section{Ensemble Equivalence at the Thermodynamic Level}
\label{section:thermo}

The thermodynamic level of ensemble equivalence
is formulated in terms of the Legendre-Fenchel transform
for concave, upper semicontinuous functions.
Such transforms arise in a natural way
via the variational formula (\ref{eqn:varphi}) for
the canonical free energy $\varphi$.  
Replacing the infimum over $y \in {\cal X}$ by
the infimum over $y \in {\cal X}$ satisfying
$\tilde{H}(y) = u$ followed by the infimum over
$u \in \rsigma$ and using the definition (\ref{eqn:entropy})
of the microcanonical entropy $s$,
we see that for all $\beta \in \rsigma$
\beas
\varphi(\beta) \! & = & \! \inf_{u \in \rsigma} \{\lan \beta,u \ran +
\inf\{I(y): y \in {\cal X}, \tilde{H}(y) = u\} \} \\
\! & = & \! \inf_{u \in \rsigma} \{\lan \beta,u \ran - s(u)\}
\, = \, s^*(\beta).
\eeas

This calculation shows that $\varphi$, the basic thermodynamic
function in the canonical ensemble, can always be expressed as
the Legendre-Fenchel transform $s^*$ of $s$, 
the basic thermodynamic function in the microcanonical
ensemble.  However, the converse need not be true.  In fact,
by the theory of Legendre-Fenchel transforms
$s(u) = \varphi^*(u)$ for all $u \in \rsigma$, or equivalently 
$s(u) = s^{**}(u)$ for all $u$,
if and only if $s$ is concave and upper semicontinuous 
on $\rsigma$.  While the upper semicontinuity is automatic
from the definition of $s$, the concavity does 
not hold in general.   This state of affairs
concerning $\varphi$ and $s$ makes it clear that the
thermodynamic level reveals what we have already
seen at the level of equilibrium macrostates; namely, 
of the two ensembles the microcanonical ensemble is the
more fundamental.

Similar considerations apply to the relationship between
$s$ and $\varphi_g$, the generalized canonical free energy, defined
in terms of a continuous function $g$ mapping $\rsigma$ into $\R$.
Making the same changes in the variational formula (\ref{eqn:varphig})
for $\varphi_g$ as we just did in the variational formula for $\varphi$
shows that for all $\beta \in \rsigma$
\beas
\varphi_g(\beta) \! & = & \! \inf_{u \in \rsigma} \{\lan \beta,u \ran +
g(u) + \inf\{I(y): y \in \rsigma, \tilde{H}(y) = u\} \} \\
\! & = & \! \inf_{u \in \rsigma} \{\lan \beta,u \ran + g(u) - s(u)\} \\
& = & (s-g)^*(\beta).
\eeas
As in the case when $g \equiv 0$, this relationship can be inverted
to give $(s-g)(u) = \varphi_g^*(u)$ for all $u \in \rsigma$,
or equivalently $(s-g)(u) = (s-g)^{**}(u)$, if
and only if $s-g$ is concave on $\rsigma$.

In order to be able to express these relationships in forms similar to those 
relating $\varphi$ and $s$, we define for $\beta$ and $u$ in $\rsigma$
\be
\label{eqn:ssharp}
s^\sharp(g,\beta) = \inf_{u \in \rsigma}\{\lan \beta,u \ran + g(u) - s(u)\} =
(s-g)^*(\beta)
\ee
and 
\be
\label{eqn:ssharpsharp}
s^{\sharp \sharp}(g,u) = g(u) + \inf_{\beta \in \rsigma}\{\lan \beta,u \ran - s^\sharp(g,\beta)\} =
g(u) + (s-g)^{**}(u).
\ee
Thus for all $\beta$, $\varphi_g(\beta) = s^\sharp(g,\beta)$ while
for all $u$, $s^{\sharp \sharp}(g,u) = s(u)$ if and only if 
$(s-g)(u) = (s-g)^{**}(u) = \varphi_g^*(u)$, and this holds if and only if $s-g$ is concave on $\rsigma$.

The next theorem records these facts in parts (a) and (b).
Part (c) introduces a new theme proved 
in Theorem 26.3 in \cite{Rock}.  The strict concavity of $s-g$ on
$\mboxdoms$ implies that $\varphi_g$ is essentially smooth; i.e., $\varphi_g$ is
differentiable on $\rsigma$ and 
\[
\lim_{n \goto \infty} \|\nabla \varphi_g(\beta_n)\|
= \infty \mbox{ whenever } \|\beta_n\| \goto \infty.
\]  
Setting $g \equiv 0$ implies a similar result relating $s$ and $\varphi_0 = \varphi$.
The differentiability of $\varphi(\beta)$ or
$\varphi_g(\beta)$ implies that the corresponding ensemble does not
exhibit a discontinuous, first-order phase transition.  

\begin{thm} \per
\label{thm:properties}
Let $g$ be a continuous function mapping $\rsigma$ into $\R$, in terms of which
the generalized canonical ensemble {\em (\ref{eqn:gencanon})}
is defined.  The choice $g \equiv 0$ gives
the standard canonical ensemble {\em (\ref{eqn:canonens})}.
The following conclusions hold.

{\em (a)} For all $\beta \in \rsigma$, $\varphi_g(\beta) = s^\sharp(g,\beta) = (s-g)^*(\beta)$.

{\em (b)}  For all $u \in \rsigma$
\[
s(u) = g(u) + (s-g)^{**}(g,u) = g(u) + \varphi_g^*(u)
\] 
if and only if $s-g$ is concave on $\rsigma$.
Both of these are equivalent to $(s-g)(u) = (s-g)^{**}(u)$
and to $s(u) = s^{\sharp\sharp}(g,u)$.

{\em (c)} If $\mboxemdoms$ is convex and $s-g$ is strictly concave on 
$\mboxdoms$, then $\varphi_g$ is essentially smooth; in particular,
$\varphi_g$ is differentiable on $\rsigma$.
\end{thm}

Theorem \ref{thm:properties} is the basis for defining 
equivalence and nonequivalence of ensembles at the thermodynamic level.
The microcanonical
and canonical ensembles are said to be thermodynamically
equivalent at $u \in \mboxdoms$ if $s(u) = s^{**}(u)$ 
and to be thermodynamically nonequivalent at $u$ if $s(u) \not = s^{**}(u)$;
the latter inequality holds if and only if $s(u) < s^{**}(u)$ [Prop.\ \ref{prop:susu}].  
Similarly, the microcanonical and
generalized canonical ensembles are said to be thermodynamically equivalent
at $u$ if $(s-g)(u) = (s-g)^{**}(u)$ --- equivalently, 
$s(u) = s^{\sharp\sharp}(g,u)$ ---
and to be thermodynamically nonequivalent at $u$ if $(s-g)(u) \leq (s-g)^{**}(u)$;
the latter inequality holds if and only if $(s-g)(u) < (s-g)^{**}(u)$ [Prop.\ \ref{prop:susu}]. 

The relationship between ensemble equivalence at the 
thermodynamic level and at the level of equilibrium macrostates is formulated
in the next theorem for the microcanonical and generalized canonical ensembles.  Setting 
$g \equiv 0$ gives the corresponding relationships between ensemble equivalence
at the two levels for the microcanonical and canonical ensembles.  Ensemble equivalence at the 
thermodynamic level involves concavity properties of $s-g$ while ensemble
equivalence at the level of equilibrium macrostates involves support properties
of $s-g$.  Except possibly for relative boundary points, $s-g$ is concave
at $u \in \mboxdoms$ if and only if $s-g$ has a supporting hyperplane at
$u$.  Hence if $\mboxdoms$ is open and so contains no relative boundary points,
then the relationship between the two levels of ensemble equivalence is
elegantly symmetric.  This is given in part (a).  In part (b) we state
the less symmetric relationship between the two levels when $\mboxdoms$ is
not open and so contains relative boundary points.

\begin{thm}\per
\label{thm:twolevels}
Let $g$ be a continuous function mapping $\rsigma$ into $\R$, in terms of which
the generalized canonical ensemble {\em (\ref{eqn:gencanon})} is defined.  
The choice $g \equiv 0$ gives the standard canonical ensemble.  The following conclusions hold.

{\em (a)} Assume that $\mbox{{\em dom}} \, s$ is an open subset of $\rsigma$.
Then the microncanonical and generalized canonical
ensembles are thermodynamically equivalent at $u \in
\mbox{\em{dom}} \, s$ if and only if the ensembles are either fully or partially equivalent
at $u$.  

{\em (b)} Assume that $\mboxemdoms$ is not an open subset of $\rsigma$.
If the microcanonical
and generalized canonical ensembles are thermodynamically equivalent at $u \in 
\mbox{\em{ri(dom}} \, s)$, then the ensembles are either fully or partially equivalent
at $u$.  Conversely, if the ensembles are either fully or partially equivalent
at $u \in \mbox{{\em dom}} \, s$, then the ensembles are thermodynamically equivalent at $u$.
\end{thm}

\noi
{\bf Proof.}  (a) If $\mboxdoms$ is open, then since $\mboxdoms$ contains no relative boundary points,
the sets $\mboxdoms$ and $\mboxridoms$ coincide.  Hence part (a) is a consequence of part (b).

(b)  If the ensembles are thermodynamically equivalent at $u \in \mboxridoms$,
then $(s-g)(u) = (s-g)^{**}(u)$.  Applying
the first inclusion in part (b) of Theorem \ref{thm:5.7} to $f = s-g$, we conclude 
the existence of $\beta$ 
such that $s$ has a supporting hyperplane at $u$ with normal vector $[\beta,-1]$.  
Parts (a) and (b) of Theorem \ref{thm:main} then imply
that the ensembles are either fully or partially equivalent at $u$.  Conversely,
if the ensembles are either fully or partially equivalent at $u \in \mboxdoms$, then 
by parts (a) and (b) of Theorem \ref{thm:main} there exists $\beta$ 
such that $s$ has a supporting hyperplane at $u$ with normal vector $[\beta,-1]$.
Applying part (a) of Theorem \ref{prop:2.9} to $f = (s-g)$, 
we conclude that $(s-g)(u) = (s-g)^{**}(u)$; i.e.,
the ensembles are thermodynamically equivalent at $u$.  This completes the proof.  \ \ink

\skp
In the next section we isolate a number of scenarios arising in applications
for which the microcanonical and generalized canonical ensembles are universally
equivalent.  This rests mainly on part (b) of Theorem \ref{thm:sgstrictlyconcave}, which
states that universal equivalence of ensembles holds if we can find
a $g$ such that $s-g$ is strictly concave on $\mboxridoms$.

\section{Universal Equivalence via the Generalized Canonical Ensemble}
\setcounter{equation}{0}
\label{section:gencanon}

This section addresses a basic foundational issue in statistical mechanics.  
In Theorems \ref{thm:applysigma1}--\ref{thm:localapply}, we show
that when the standard canonical ensemble is nonequivalent to the microcanonical ensemble
on a subset of values of $u$, 
it can often be replaced by a generalized canonical ensemble that is
univerally equivalent to the microcanonical ensemble.  
In three of these four theorems, the function
$g$ defining the generalized canonical ensemble is a quadratic function, and the
ensemble is Gaussian.  

In these three theorems our strategy is to find a quadratic function $g$ such
that $s-g$ is strictly concave on $\mboxridoms$ and continuous on $\mboxdoms$.
Part (b) of Theorem \ref{thm:sgstrictlyconcave} then yields the universal equivalence.  As the next 
proposition shows, an advantage of working with quadratic functions is that
support properties of $s-g$ involving a supporting hyperplane are equivalent
to support properties of $s$ involving a supporting paraboloid defined in
terms of $g$.  This observation gives a geometrically intuitive way
to find a quadratic function $g$ guaranteeing universal ensemble equivalence.

In order to state the proposition, we need a definition.
Let $f$ be a function mapping $\rsigma$ into $\R \cup \{-\infty\}$,
$u$ and $\beta$ points in $\rsigma$, and $\gamma \geq 0$.  
We say that $f$ has a supporting
paraboloid at $u \in \R^\sigma$ with parameters $(\beta,\gamma)$ if 
\[
f(v) \leq  f(u) + \langle \beta, v-u \rangle + \gamma\|v-u\|^2 \ \mbox{ for all } v \in \R^\sigma.
\]
The paraboloid is said to be strictly supporting
if the inequality is strict for all $v \not = u$.

\begin{prop}\per
\label{prop:paraboloid}
$f$ has a {\em (}strictly{\em )} 
supporting paraboloid at $u$ with parameters $(\beta,\gamma)$ if and
only if $f - \gamma \|\cdot\|^2$ has a {\em (}strictly{\em )} supporting hyperplane at $u$ with normal vector 
$[\tilde\beta,-1]$.  The quantities 
$\beta$ and $\tilde{\beta}$ are related by $\tilde{\beta} = \beta
- 2\gamma u$.
\end{prop}

\noi
{\bf Proof.}  The proof is based on the identity 
$\|v-u\|^2 = \|v\|^2 - 2\lan u, v - u \ran - \|u\|^2$.
If $f$ has a strictly supporting
paraboloid at $u$ with parameters $(\beta,\gamma)$,
then for all $v \not = u$  
\[
f(v) - \gamma\|v\|^2 < f(u) - \gamma\|u\|^2 + \lan \tilde{\beta}, v - u \ran,
\]
where $\tilde{\beta} = \beta - 2\gamma u$.  Thus $f - \gamma \|\cdot\|^2$ has a
strictly supporting
hyperplane at $u$ with normal vector $[\tilde{\beta},-1]$.  The converse is proved 
similarly, as is the case in which the supporting hyperplane or paraboloid
is supporting but not strictly supporting.  \ \ink

\skp
The first application of Theorem \ref{thm:sgstrictlyconcave}
is Theorem \ref{thm:applysigma1}, which is formulated for dimension $\sigma = 1$.
The theorem gives a criterion guaranteeing the existence of a
quadratic function $g$ such that $s-g$ is
strictly concave on $\mboxdoms$.
The criterion --- that 
$s''$ is bounded above on the interior of $\mboxdoms$ --- is essentially
optimal for the existence of a fixed
quadratic function $g$ guaranteeing the strict concavity of $s - g$. 
The situation
in which $s''$ is not bounded above on the interior of $\mboxdoms$
can often be handled by Theorem \ref{thm:localapply}, which is a local version of
Theorem \ref{thm:applysigma1}.   

The strict concavity of $s-g$ on $\mboxdoms$  has several important consequences
concerning universal equivalence of ensembles at the level
of equilibrium macrostates and equivalence of ensembles at the thermodynamic
level --- i.e., $s^{\sharp \sharp}(g,u) = s(u)$ for all $u$.  As we note
in part (e) of Theorem \ref{thm:applysigma1},
the strict concavity of $s-g$ also implies that the generalized canonical free energy
$\varphi_g = (s-g)^*$ is differentiable on $\r$,
a condition guaranteeing the absence of a discontinuous, first-order phase
transition with respect to the Gaussian ensemble.

Theorem \ref{thm:apply} is the analogue of Theorem \ref{thm:applysigma1}
that treats arbitrary dimension $\sigma \geq 2$.
When $\sigma \geq 2$, in general the results are weaker than when $\sigma = 1$.  

\begin{thm}\per
\label{thm:applysigma1}
Assume that the dimension $\sigma = 1$
and that $\mboxdoms$ is a nonempty interval.
Assume also that $s$ is continuous on $\mboxemdoms$, $s$ is twice continuously differentiable
on $\mboxemintdoms$, and $s''$ 
is bounded above on $\mboxemintdoms$.  
Then for all sufficiently large $\gamma \geq 0$ and $g(u) = \gamma u^2$,
conclusions {\em (a)--(e)} hold.  Specifically, if $s$ is strictly concave
on $\mboxemdoms$, then we choose any $\gamma \geq 0$, and otherwise we choose
\be
\label{eqn:gamma0sigma1}
\gamma > \gamma_0 = \textstyle \frac{1}{2} \cdot 
\displaystyle \sup_{u \in \mbox{\scriptsize {\em int(dom}} \, s)} s''(u).
\ee

{\em (a)}  $s-g$ is strictly concave and continuous on $\mboxemdoms$.

{\em (b)} $s-g$ has a 
strictly supporting line, and $s$ has a strictly supporting 
paraboloid, at all $u \in \mboxemdoms$ except possibly
boundary points.  At a boundary point $s-g$ has a strictly supporting line, and $s$
has a strictly supporting parabola, 
if and only if the one-sided derivative of $s-g$
is finite at that boundary point.

{\em (c)} The microcanonical ensemble and the Gaussian
ensemble defined in terms of this $g$ are universally equivalent; 
i.e., fully equivalent at all $u \in \mboxemdoms$ except possibly boundary points.
For all $u \in \mboxemintdoms$ the value of $\beta$ defining the universally
equivalent Gaussian ensemble is 
unique and is given by $\beta = s'(u) - 2\gamma u$. 

{\em (d)} For all $u \in \r$, $s^{\sharp \sharp}(g,u) = s(u)$ or equivalently
$(s-g)^{**}(u) = (s-g)(u)$.

{\em (e)} The generalized canonical free energy $\varphi_g =
(s-g)^*$ is essentially smooth;
in particular, $\varphi_g$ is differentiable
on $\rsigma$.

\end{thm}

\noi
{\bf Proof.}
(a) If $s$ is strictly concave on $\mboxdoms$, then
$s(u) - \gamma u^2$ is also strictly concave on this set for 
any $\gamma \geq 0$.  We now consider the case in which $s$ is not strictly concave 
on $\mboxdoms$.  If $g(u) = \gamma u^2$, then $s-g$ is continuous on $\mboxdoms$.  
If, in addition, we choose $\gamma > \gamma_0$ in accordance with (\ref{eqn:gamma0sigma1}),
then for all $u \in \mboxintdoms$
\[
(s-g)''(u) = s''(u) - 2 \gamma < 0.
\]
A straightforward extension of the proof of 
Theorem 4.4 in \cite{Rock}, in which the inequalities
in the first two displays are replaced by strict inequalities,
shows that $-(s-g)$ is strictly convex on $\mboxintdoms$ and thus
that $s-g$ is strictly concave on $\mboxintdoms$. 
If $s-g$ is not strictly concave on $\mboxdoms$, 
then $s-g$ must be affine on an interval.  Since this
violates the strict concavity on $\mboxintdoms$, part (a) is proved.

(b) The first assertion follows from part (a) of the present theorem,
part (a) of Theorem \ref{thm:sgstrictlyconcave}, and Proposition \ref{prop:paraboloid}.
Concerning the second assertion about boundary points, the reader is
referred to the discussion before Theorem \ref{thm:strictlyconcave}.

(c) The universal equivalence of the two ensembles is a consequence
of part (a) of the present theorem and part (b) of Theorem \ref{thm:sgstrictlyconcave}.
The full equivalence of the ensembles at all $u \in \mboxintdoms$
is equivalent to the existence of a strictly
supporting hyperplane at all $u \in \mboxintdoms$ with supergradient $\beta$ [Thm.\ \ref{thm:main}(a)].
Since $s(u) - \gamma u^2$ is differentiable at all $u \in \mboxintdoms$, part (b) of
Theorem \ref{thm:usefulfacts} implies that $\beta$ is unique
and $\beta = (s(u) - \gamma u^2)'$.

(d) The strict concavity of $s - g$ on $\mboxdoms$ proved in part (a)
implies that $s-g$ is concave on $\r$.  Part (b) of Theorem \ref{thm:properties} 
allows us to conclude that for all $u \in \r$, $s^{\sharp \sharp}(g,u)=s(u)$ 
or equivalently $(s-g)^{**}(u) = (s-g)(u)$.

(e) This follows from part (c) of Theorem \ref{thm:properties}. \ \ink

\skp

We now consider the analogue of Theorem \ref{thm:applysigma1} for arbitrary
dimension $\sigma \geq 2$.   In contrast to the case
$\sigma = 1$, in which $s-g$ could always be extended to a strictly concave
function on all of $\mboxdoms$, in the case $\sigma \geq 2$
there exists a quadratic $g$ such that $s-g$ is strictly concave on the
interior of $\mboxdoms$, but in general $s-g$ cannot be extended to a strictly concave 
function on all of $\mboxdoms$.  
One can easily find examples in which the boundary of $\mboxdoms$ has
flat portions and $s-g$ is strictly concave on the interior of
$\mboxdoms$ and constant on these flat portions.
As a result, unless $\mboxdoms$ is open,
we cannot apply part (c) of Theorem \ref{thm:properties} to conclude
that the generalized canonical free energy
$\varphi_g = (s-g)^*$ is differentiable on $\rsigma$. 

\begin{thm}\per
\label{thm:apply}
Assume that the dimension $\sigma \geq 2$ and 
that $\mboxemdoms$ is convex and has nonempty interior.
Assume also that $s$ is continuous on $\mboxemdoms$, $s$ is twice continuously differentiable
on $\mboxemintdoms$, and 
all second-order partial derivatives of $s$
are bounded above on $\mboxemintdoms$.  
Then for all sufficiently large $\gamma \geq 0$ and $g(u) = \gamma \|u\|^2$, 
conclusions {\em (a)--(e)} hold.  Specifically, if $s$ is strictly
concave on $\mboxemintdoms$, then we choose any $\gamma \geq 0$, and otherwise
we choose
\be
\label{eqn:gamma0}
\gamma > \gamma_0 = \textstyle \frac{1}{2} \cdot 
\displaystyle \sup_{u \in \mbox{\scriptsize {\em int(dom}} \, s)} \kappa(u),
\ee
where $\kappa(u)$ denotes the largest eigenvalue of the symmetric Hessian matrix
of $s$ at $u$.  

{\em (a)}  $s-g$ is strictly concave on $\mboxemintdoms$ and 
concave and continuous on $\mboxdoms$.

{\em (b)} $s-g$ has a 
strictly supporting hyperplane, and $s$ has a strictly supporting
paraboloid, at all $u \in \mboxemdoms$ except
possibly boundary points.

{\em (c)} The microcanonical ensemble and the Gaussian
ensemble defined in terms of this $g$ are universally equivalent; 
i.e., fully equivalent at all $u \in \mboxemdoms$ except possibly boundary points.
For all $u \in \mboxemintdoms$ the value of $\beta$ defining 
the universally equivalent Gaussian ensemble is unique and is given by
$\beta = \nabla s(u) - 2\gamma u$.  

{\em (d)} For all $u \in \rsigma$, $s^{\sharp \sharp}(g,u) = s(u)$ or equivalently
$(s-g)^{**}(u)= (s-g)(u)$. 

{\em (e)} Assume that $\mbox{{\em dom}} \, s$ is open.  Then 
the generalized canonical free energy $\varphi_g =
(s-g)^*$ is essentially smooth; 
in particular, $\varphi_g$ is differentiable on $\rsigma$.
\end{thm}

\noi
{\bf Proof.}  
(a) If $s$ is strictly concave on $\mboxintdoms$, then 
$s - \gamma \|\cdot\|^2$ is also strictly concave on this set for any $\gamma \geq 0$.
We now consider the case in which $s$ is not strictly concave on $\mboxintdoms$.
If $g(u) = \gamma \|u\|^2$, then $s-g$ is continuous on $\mboxdoms$.  
For $u \in \mboxintdoms$, let 
$Q_u = \{\partial\,^2 s(u)/\partial u_i \partial u_j\}$ denote the Hessian matrix
of $s$ at $u$.  We choose $\gamma > \gamma_0$ in accordance with (\ref{eqn:gamma0}),
noting that 
\bea
\label{eqn:eigenvalue}
\gamma_0 & = & \textstyle \frac{1}{2} \cdot \displaystyle 
\sup_{u \in \mbox{\scriptsize int(dom} \, s)} \kappa(u) \\
 & = & \textstyle \frac{1}{2} \cdot \displaystyle \sup_{u \in \mbox{\scriptsize int(dom} \, s)}
\sup\!\left\{\lan Q_u \zeta,\zeta \ran : \zeta \in \rsigma, \|\zeta\| = 1\right\}.
\nonumber
\eea
Let $I$ be the identity matrix.  It follows that for any $u \in \mboxintdoms$
and all nonzero $z \in \rsigma$
\[
\lan (Q_u - 2 \gamma I)z,z \ran < 0.
\]
By analogy with the proof of Theorem 4.5 in \cite{Rock}, 
the strict concavity of $s-g$ on $\mboxintdoms$
is equivalent to the strict concavity of
$s-g$ on each line segment in $\mboxintdoms$.  This, in turn,
is equivalent to the strict concavity, for each
$v \in \mboxintdoms$ and nonzero $z \in \rsigma$, of $\psi(\lambda)
= (s-g)(v + \lambda z)$ on the open interval
$G(v,z) = \{\lambda \in \R : v + \lambda z \in \mboxintdoms\}$.
Since
\[
\psi''(\lambda) = \lan (Q_{v + \lambda z} - 2\gamma I)z,z \ran < 0,
\]
$\psi'$ is strictly decreasing on $G(v,z)$.
A straightforward extension of the proof of 
Theorem 4.4 in \cite{Rock}, in which the inequalities
in the first two displays are replaced by strict inequalities,
shows that $-\psi$ is strictly convex on $G(v,z)$ and thus
that $\psi$ is strictly concave on $G(v,z)$.  It follows
that $s-g$ is strictly concave on $\mboxintdoms$.  
By Proposition \ref{prop:extend} the continuity of $s-g$ on $\mboxdoms$
allows us to extend the strict concavity of $s-g$ on $\mboxintdoms$ 
to the concavity of $s-g$ on $\mboxdoms$.  This completes the proof of part (a).

(b)--(d)  These are proved as in Theorem \ref{thm:applysigma1}. 

(e) If $\mbox{dom} \, s$ is open, then
part (a) implies that $s - g$ is strictly concave on 
$\mbox{dom} \, s$.
The essential smoothness of $(s-g)^*$, and thus its differentiability,
are consequences of part (c) of Theorem \ref{thm:properties}. \ \ink

\skp

In the next theorem we give other conditions on $s$ guaranteeing conclusions
similar to those in Theorems \ref{thm:applysigma1} and \ref{thm:apply}.

\begin{thm}\per
\label{thm:apply2} Assume that $\mbox{{\em dom}} \, s$ is convex, closed, and bounded and 
that $s$ is bounded and continuous on $\mbox{{\em dom}} \, s$.   Then there 
exists a continuous function $g$ mapping $\rsigma$ into $\R$ such
that the following conclusions hold.

{\em (a)} $s - g$ is strictly concave and continuous on $\mbox{{\em dom}} \, s$,
and the generalized canonical free energy $\varphi_g =
(s-g)^*$ is essentially smooth; in particular,
$\varphi_g$ is differentiable
on $\rsigma$.

{\em (b)} $s-g$ has a strictly supporting hyperplane at all 
$u \in \mbox{{\em dom}} \, s$ except possibly relative boundary points.

{\em (c)} The microcanonical ensemble and the generalized
canonical ensemble defined in terms of this $g$ are universally equivalent
on $\mbox{{\em dom}} \, s$; i.e., fully equivalent at 
all $u \in \mbox{{\em dom}} \, s$ except possibly relative boundary points.

{\em (d)} For all $u \in \rsigma$, $s^{\sharp \sharp}(g,u) = s(u)$ or equivalently
$(s-g)^{**}(u) = (s-g)(u)$.
\end{thm}

\noi
{\bf Proof.}  (a) Let $h$ be any strictly concave function on $\rsigma$.   
Since $h$ is continuous
on $\rsigma$ \cite[Cor.\ 10.1.1]{Rock}, $h$ is also bounded and continuous on $\mboxdoms$ . 
For $u \in \mbox{dom} \, s$ define $g(u) = s(u) - h(u)$.
Since $g$ is bounded and continuous on the closed set $\mbox{dom} \, s$, 
the Tietze Extension Theorem guarantees that $g$ can be extended to a bounded, continuous function
on $\rsigma$ \cite[Thm.\ 4.16]{Fol}.  Then $s - g$
has the properties in part (a).  The strict concavity of $s - g$ on $\mbox{dom} \, s$
implies the essential smoothness of $(s-g)^*$ and thus its differentiability 
[Thm.\ \ref{thm:properties}(c)].  

(b) This follows from part (a) of the present theorem
and part (a) of Theorem \ref{thm:sgstrictlyconcave}.

(c) The universal equivalence of the two ensembles is a consequence
of part (a) of the present theorem and part (b) of Theorem \ref{thm:sgstrictlyconcave}.

(d) The function $g$ constructed in the proof of part (a) is bounded and 
continuous on $\rsigma$.  In addition, $s - g$ is strictly concave
on $\mbox{dom} \, s$ and thus concave on $\rsigma$.  
Since $s-g$ is continuous on the closed set
$\mboxdoms$, $s-g$ is also upper semicontinuous on $\rsigma$.  
Part (b) of Theorem \ref{thm:properties} implies that
for all $u \in \rsigma$, $s^{\sharp \sharp}(u)=s(u)$ or equivalently
$(s-g)^{**}(u) = (s-g)(u)$.
\ \ink

\skp
Suppose that $s$ is $C^2$ on the interior of $\mboxdoms$ but
the second-order partial derivatives of $s$ are not bounded above.
This arises, for example, in the Curie-Weiss-Potts model, in which
$\mbox{dom} \, s$ is a closed, bounded interval of $\R$ and
$s''(u) \goto \infty$ as $u$ approaches the right hand endpoint of
$\mbox{dom} \, s$ \cite{CosEllTou1}.
In such cases one cannot expect that the conclusions of Theorems
\ref{thm:applysigma1} and \ref{thm:apply} will be satisfied; 
in particular, that there exists a quadratic function $g$
such that $s-g$ has a strictly
supporting hyperplane at each point of the interior of $\mboxdoms$ and thus that
the ensembles are universally equivalent.

In order to overcome this difficulty,
we introduce Theorem \ref{thm:localapply}, a local version of Theorems \ref{thm:applysigma1}
and \ref{thm:apply}.  Theorem \ref{thm:localapply} handles the case in which $s$ is $C^2$
on an open set $K$ but either $K$ is not all of $\mboxintdoms$
or $K = \mboxintdoms$ and the second-order partial
derivatives of $s$ are not all bounded above on $K$.
In neither of these situations are the hypotheses of Theorem \ref{thm:applysigma1}
or \ref{thm:apply} satisfied.
In Theorem \ref{thm:localapply} additional conditions are given 
guaranteeing that for each $u \in K$ 
there exists $\gamma$ depending on $u$ such that
$s - \gamma\|\cdot\|^2$ has a strictly supporting hyperplane at $u$.  
Our strategy is first to choose a paraboloid that is strictly supporting in a neighborhood of $u$
and then to adjust $\gamma$ so that the paraboloid becomes
strictly supporting on all $\rsigma$.  Proposition \ref{prop:paraboloid}
then guarantees that $s - \gamma \|\cdot\|^2$ has a strictly
supporting hyperplane at $u$.   

This construction for each $u \in K$
implies a form of universal equivalence of ensembles that
is weaker than that in Theorems \ref{thm:applysigma1} and \ref{thm:apply} but is still useful.
In contrast to those theorems, which state that
$s^{\sharp \sharp}(g,u) = s(u)$ for all $u \in \rsigma$,
in Theorem \ref{thm:localapply}
we prove the alternative representation $\inf_{\gamma \geq 0}s^{\sharp \sharp}(g_\gamma,u) = s(u)$ for all
$u$ in $K$, where $g_\gamma = \gamma \|\cdot\|^2$ for $\gamma \geq 0$.  
This alternative representation is necessitated by the fact
that the quadratic depends on $u$.

For each fixed $u \in K$ the value of $\gamma$ for which $s - \gamma \|\cdot\|^2$ has a strictly
supporting hyperplane at $u$ depends on $u$.
However, with the same $\gamma$ one might also have a strictly supporting
hyperplane at other values of $u$.  In general, as one increases $\gamma$, 
the set of $u$ at which $s-\gamma\|\cdot\|^2$ has a strictly supporting hyperplane cannot decrease.
Because of part (a) of Theorem \ref{thm:main}, this can be restated in terms of
ensemble equivalence involving the Gaussian ensemble and the
corresponding set $\egammabeta$ of equilibrium macrostates defined 
in (\ref{eqn:egammabeta}).  Defining
\[
U_\gamma = \{u \in K : \mbox{there exists } \beta \mbox{ such that } \egammabeta = \eu\},
\]
we have $U_{\gamma_1} \subset U_{\gamma_2}$ whenever $\gamma_2 >
\gamma_1$ and because of Theorem \ref{thm:localapply},
$\bigcup_{\gamma > 0} U_\gamma = K$. 
This phenomenon is investigated in detail in \cite{CosEllTou2} for the Curie-Weiss-Potts model.

In order to state Theorem \ref{thm:localapply}, we define
for $u \in K$ and $\lambda \geq 0$ 
\[
\label{dx0s}
D(u,\nabla s(u), \lambda) = \left\{v \in \mboxdoms : s(v) \geq s(u) + \langle \nabla s(u),
v - u \rangle + \lambda \|v - u\|^2 \right\}.
\]
Geometrically, this set contains all points for which the 
paraboloid with parameters $(\nabla s(u),\lambda)$ passing through $(u,s(u))$
lies below the graph of $s$.
Clearly, since $\lambda \geq 0$, we have
$D(u,\nabla s(u), \lambda)$ $\subset D(u,\nabla s(u),0)$;
the set $D(u,\nabla s(u),0)$ contains
all points for which the 
graph of the hyperplane with normal vector $[\nabla s(u),-1]$ 
passing through $(u,s(u))$ lies below the 
graph of $s$.  Thus, in the next theorem
the hypothesis that for each $u \in K$ the set 
$D(u,\nabla s(u), \lambda)$ is bounded for some $\lambda \geq 0$
is satisfied if $\mboxdoms$ is bounded or, more generally, if
$D(s,\nabla s(u),0)$ is bounded.  
The latter set is bounded if, for example, $-s$ is superlinear; 
i.e., \[
\lim_{\|v\| \goto \infty} s(v)/\|v\| = -\infty.
\]
As we have remarked, the next theorem can often be applied when the
hypotheses of Theorem \ref{thm:applysigma1} or \ref{thm:apply}
are not satisfied.

\begin{thm}\per
\label{thm:localapply}
Let $K$ an open subset of $\mbox{{\em dom}} \, s$ and 
assume that $s$ is twice continuously differentiable on $K$.
Assume also that $\mboxemdoms$ is bounded or, more generally, that
for every $u \in \mbox{{\em int}} \, K$ there exists $\lambda \geq 0$ such that
$D(u,\nabla s(u), \lambda)$ is bounded.
The following conclusions hold.

{\em (a)} For each $u \in K$, define $\gamma_0(u) \geq 0$ by {\em (\ref{eqn:gamma0A})}.
Then for any $\gamma > \gamma_0(u)$, $s$ has a strictly supporting
paraboloid at $u$ with parameters $(\nabla s(u),\gamma)$. 

{\em (b)} For each $u \in K$ we choose $\gamma > \gamma_0(u)$
as in part {\em (a)} and define $g_\gamma = \gamma\|\cdot\|^2$.
Then $s - g_\gamma$ has a strictly supporting hyperplane at $u$ with
normal vector $[\nabla s(u) - 2 \gamma u,-1]$.

{\em (c)}  For each $u \in K$
\[
\inf_{\gamma \geq 0}s^{\sharp \sharp}(g_\gamma,u) = 
\inf_{\gamma \geq 0}\{g_\gamma(u) + (s-g_\gamma)^{**}(u)\} = s(u).
\]

{\em (d)} For each $u \in K$
choose $g = \gamma \|\cdot\|^2$ such that, in accordance with
part {\em (b)}, $s-g$ has a strictly supporting hyperplane
at $u$.  Then the microcanonical ensemble and the Gaussian
ensemble defined in terms of this $g$ are fully equivalent at $u$.  
The value of $\beta$ defining the Gaussian ensemble is unique and is given by
$\beta = \nabla s(u) - 2\gamma u$.  
\end{thm}

\noi
{\bf Proof.} (a)  
Given $u \in K$, let $B(u,r) \subset K$
be an open ball with center $u$ and positive radius $r$ whose closure
is contained in $K$.  If the dimension $\sigma = 1$, then $s''$ is bounded above
on $B(u,r)$, while if $\sigma \geq 2$, then all second-order
partial derivatives of $s$ are bounded above on $B(u,r)$.
We now apply, to the restriction of $s$ to $B(u,r)$, part (a) of
Theorem \ref{thm:applysigma1} when $\sigma = 1$ and
part (a) of Theorem \ref{thm:apply} when $\sigma \geq 2$.
We conclude that there exists a
sufficiently large $A \geq 0$ such that $s - A\|\cdot\|^2$ is strictly
concave on $B(u,r)$.  Part (c) of Theorem \ref{prop:2.9} 
implies that when restricted to $B(u,r)$,
$s - A\|\cdot\|^2$ has a strictly supporting
hyperplane at $u$; that is, there exists $\theta \in
\R^\sigma$ such that 
\be
\label{eqn:fminusa}
s(v) - A\|v\|^2 < s(u) -A\|u\|^2 + \langle \theta, v-u \rangle
\ \mbox{ for all } v \in B(u,r), v \not = u.
\ee
In fact, $\theta = \nabla s(u) - 2Au$ because $s - A\|\cdot\|^2$ is concave
and differentiable on $B(u,r)$ [Thm.\ \ref{thm:usefulfacts}(b)].  
We rewrite the inequality in
the last display as 
\be
\label{eqn:lastdisplay}
s(v) < s(u) + \langle \nabla s(u), v-u \rangle + A\|v-u\|^2 
\ \mbox{ for all } v \in B(u,r), v \not = u.
\ee
This inequality continues to hold if we take larger values of $A$, and so
without loss of generality we can assume that $A > \lambda$.  
Because $s(v) = -\infty$ for $v \notin \mboxdoms$,
the set where the inequality in the last display does not hold
is $D(u,\nabla s(u), A)$.  
Since $A > \lambda$, we have $D(u,\nabla s(u), A) \subset
D(u,\nabla s(u), \lambda)$, and since the latter
set is assumed to be bounded, 
there exists $b \in (0,\infty)$ such that 
\be
\label{eqn:dx0nabla}
D(u,\nabla s(u), A) \subset \{v \in \rsigma : \|v-u\| < b \}.
\ee

Let $\gamma$ be any number satisfying
\be
\label{eqn:gamma0A}
\gamma > \gamma_0(u) = \max\!\left\{A, \frac{-s(u) + \|\nabla s(u)\|b}{r^2} \right\}.
\ee
Since $A \geq 0$, it follows that $\gamma_0(u) \geq 0$.
We now prove that $s$ has a strictly supporting paraboloid at $u$ 
with parameters $(\nabla s(u),\gamma)$; i.e., 
\be
\label{eqn:whatwewant}
s(v) < s(u) + \langle \nabla s(u), v-u \rangle + \gamma\|v-u\|^2
\ \mbox{ for all } v \in \rsigma, v \not = u.
\ee
It suffices to prove (\ref{eqn:whatwewant}) for all $v \in \mboxdoms$.
Since $\gamma > A$ and since (\ref{eqn:lastdisplay}) is valid for 
all $v \in B(u,r)$, $v \not = u$, (\ref{eqn:whatwewant}) is also 
valid for all $v \in B(u,r)$, $v \not = u$.  In addition, for all 
$v \in \mboxdoms \setminus D(u,\nabla s(u), A)$
\beas
s(v) & < & s(u) + \langle \nabla s(u), v-u \rangle + A\|v-u\|^2 \\
& \leq & s(u) + \langle \nabla s(u), v-u \rangle + \gamma\|v-u\|^2,
\eeas
and so (\ref{eqn:whatwewant}) is also valid for all such $v$. 
We finally show that (\ref{eqn:whatwewant})
is valid for all 
$v \in D(u,\nabla s(u), A) \setminus B(u,r)$.
This follows from the string of inequalities 
\beas
\lefteqn{
s(u) + \langle \nabla s(u), v-u \rangle + \gamma\|v-u\|^2 } \\
& & \hspace{.4in} > s(u) + \langle \nabla s(u), v-u \rangle + \gamma r^2 \\
& & \hspace{.4in} >  s(u) - \|\nabla s(u)\|b -s(u) + \|\nabla s(u)\|b \\
& & \hspace{.4in} = 0 \\
&& \hspace{.4in} \geq s(v).
\eeas
By proving that (\ref{eqn:whatwewant}) is valid for all $v \in \rsigma$,
we have completed the proof of part (a).

(b) This follows from part (a) of the present theorem and 
Proposition \ref{prop:paraboloid}.

(c) By part (b), for each $u \in K$ and
any $\tilde\gamma > \gamma_0$, $s-g_{\tilde\gamma}$ has a strictly supporting hyperplane,
and thus a supporting hyperplane, at $u$.  
We now apply to $s-g_{\tilde\gamma}$ part (a)
of Theorem \ref{prop:2.9}, obtaining 
$(s-g_{\tilde\gamma})^{**}(u) = (s-g_{\tilde\gamma})(u)$ 
or 
\[
s(u) = g_{\tilde\gamma}(u) + (s-g_{\tilde\gamma})^{**}(u).
\]  
Since for any $\gamma \geq 0$, $(s - g_{\gamma})^{**}(u) \geq
(s-g_{\gamma})(u)$ [Prop.\ \ref{prop:susu}], 
it follows from (\ref{eqn:ssharpsharp}) that 
\[
s(u) = \inf_{\gamma \geq 0}\{g_\gamma(u) + (s-g_\gamma)^{**}(u)\} = 
\inf_{\gamma \geq 0}s^{\sharp\sharp}(g,u).
\]

(d) Fix $u \in K$ and let $B(u,r)$ be an open ball with center $u$
and radius $r$ whose closure is contained in $K$.
The full equivalence of the ensembles follows from part (b) of the present
theorem and part (a) of Theorem
\ref{thm:main}.  The value of $\beta$ defining the fully equivalent
Gaussian ensemble is characterized by the property that $[\beta,-1]$ is the normal
vector to a strictly supporting hyperplane for $s-\gamma\|\cdot\|^2$ at $u$.  
In order to identify $\beta$, we consider the convex function $h$
that equals $s - \gamma \|\cdot\|^2$ on the open ball $B(u,r)$
and equals $-\infty$ on the complement.  Since $h$ is differentiable at $u$, 
part (b) of Theorem \ref{thm:usefulfacts} implies that $\beta$ is unique
and equals $\nabla h(u) = \nabla(s - \gamma \|\cdot\|^2)(u)$.
This completes the proof. \ \ink

\skp
Theorem \ref{thm:localapply} suggests an extended form of the notion of universal
equivalence of ensembles.  In Theorems \ref{thm:applysigma1}--\ref{thm:apply2} we 
are able to achieve full equivalence of ensembles for all 
$u \in \mboxdoms$ except possibly relative
boundary points by choosing an appropriate $g$ that
is valid for all $u$.  This leads to the observation
in each theorem that
the microcanonical ensemble and the generalized canonical
ensemble defined in terms of this $g$ are universally
equivalent.  In Theorem \ref{thm:localapply}
we can also achieve full equivalence of ensembles
for all $u \in K$.  However, in contrast to 
Theorems \ref{thm:applysigma1}--\ref{thm:apply2},
the choice of $g$ for which the two ensembles are
fully equivalent depends on $u$.  We summarize the ensemble equivalence
property articulated in part (d) of Theorem \ref{thm:localapply} by
saying that relative to the set of quadratic functions,
the microcanonical ensemble and the Gaussian
ensembles are universally equivalent on the open set $K$ of mean-energy values.

We complete our discussion of the generalized canonical ensemble and its
equivalence with the microcanonical ensemble by noting that 
the smoothness hypothesis on $s$ in Theorem \ref{thm:localapply}
is essentially satisfied whenever
the microcanonical ensemble exhibits no phase transition at any $u \in K$.
In order to see this, 
we recall that a point $u_c$ at which $s$ is not differentiable represents
a first-order, microcanonical phase transition \cite[Fig.\ 3]{ETT}.  
In addition, a point $u_c$ at which $s$ is differentiable
but not twice differentiable represents a second-order, microcanonical phase transition
\cite[Fig.\ 4]{ETT}.  It follows that $s$ is smooth on any open set $K$ not containing 
such phase-transition points.  Hence, if
the other conditions in Theorem \ref{thm:localapply} are valid, then 
the microcanonical and Gaussian ensembles are universally equivalent on $K$ relative
to the set of quadratic functions.  In particular, if the microcanonical ensemble
exhibits no phase transitions, then $s$ is smooth on all of $\mboxintdoms$.  This implies the
universal equivalence of the two ensembles provided that the other
conditions are valid in Theorem \ref{thm:applysigma1} if $\sigma = 1$
or in Theorem \ref{thm:apply} if $\sigma \geq 2$.

\appendix
\section{Material on Concave Functions}
\setcounter{equation}{0}

\renewcommand{\thesection}{\Alph{section}}
\renewcommand{\theequation}
{\Alph{section}.\arabic{equation}}
\renewcommand{\thedefn}
{\Alph{section}.\arabic{defn}}
\renewcommand{\theass}
{\Alph{section}.\arabic{ass}}

This appendix contains
a number of technical results on concave functions needed in the
main body of the paper. The theory of concave functions, rather than that of convex functions, is
the natural setting for statistical mechanics.  This is convincingly illustrated
by the main theme of this paper, which is that concavity and strict concavity
properties of the microcanonical entropy are closely related to the equivalence
and nonequivalence of the microcanonical and canonical ensembles.

Let $\sigma$ be a positive integer.
A function $f$ on $\rsigma$ is said to be concave on $\rsigma$, or concave, if $-f$ is a proper
convex function in the sense of \cite[p.\ 24]{Rock}; 
that is, $f$ maps $\rsigma$ into $\r \cup \{-\infty\}$, $f \not \equiv 
-\infty$, and for all $u$ and $v$ in $\rsigma$ and all $\lambda \in (0,1)$
\[
\label{eqn:concave}
f(\lambda u + (1-\lambda)v) \geq \lambda f(u) + (1-\lambda) f(v).
\]
 
Given $f \not \equiv -\infty$ a function mapping $\rsigma$ into
$\R \cup \{-\infty\}$, we define $\mboxdomf$ to be the set of
$u \in \rsigma$ for which $f(u) > -\infty$.  
Let $\beta$ be a point in $\rsigma$.
The function $f$ is said to have a supporting hyperplane at $u \in \mboxdomf$
with normal vector $[\beta,-1]$ if 
\[
\label{eqn:supphyp}
f(v) \leq  f(u) + \langle \beta, v-u \rangle \ \mbox{ for all } v \in \rsigma.
\]
It follows from this inequality that $u \in \mboxdomf$.  
In addition, $f$ is said to have a strictly supporting hyperplane 
at $u \in \mboxdomf$ with normal vector $[\beta,-1]$ if 
the inequality in the last display is strict for all $v \not = u$.  

Two useful facts for concave functions on $\rsigma$ are given in the next theorem.
They are proved in Theorems 23.4 and 25.1 in \cite{Rock}.
The quantities appearing in Theorem \ref{thm:usefulfacts} are defined after Corollary \ref{cor:third}
in the present paper.

\begin{thm}\per
\label{thm:usefulfacts}
Let $f$ be a concave function on $\rsigma$.  The following conclusions hold.

{\em (a)} $\mboxemridomf \subset \mbox{{\em dom}} \, \partial f \subset \mboxemdomf$.

{\em (b)} If $f$ is differentiable at $u \in \mboxemdomf$, then $\nabla f(u)$
is the unique supergradient of $f$ at $u$.
\end{thm}

Let $f \not \equiv -\infty$ be a function mapping $\rsigma$ into $\R \cup \{-\infty\}$.
For $\beta$ and $u$  in $\rsigma$ the Legendre-Fenchel
transforms $f^*$ and $f^{**}$ are defined by \cite[p.\ 308]{Rock}
\[
f^*(\beta) = \inf_{u \in \rsigma} \{\lan \beta,u \ran - f(u)\}
\ \mbox{ and } \
f^{**}(u) = \inf_{\beta \in \rsigma} \{\lan \beta,u \ran - f^*(\beta)\}.
\]
As in the case of convex functions \cite[Thm.\ VI.5.3]{Ell}, 
$f^*$ is concave and upper semicontinuous on $\rsigma$
and for all $u \in \rsigma$ we have $f^{**}(u) = f(u)$ if and only if 
$f$ is concave and upper semicontinuous on $\rsigma$.  When $f$ 
is not concave and upper semicontinuous, the relationship between
$f$ and $f^{**}$ is given in the next proposition.
  
\begin{prop}\per
\label{prop:susu}
Let $f \not \equiv -\infty$ be a function mapping $\rsigma$ into
$\R \cup \{-\infty\}$.  If $f$ is not concave and upper semicontinuous on $\rsigma$, then 
$f^{**}$ is the smallest concave, upper semicontinuous function on $\rsigma$
that satisfies $f^{**}(u) \geq f(u)$ for all $u \in \rsigma$.  In particular,
if for some $u$, $f(u) \not = f^{**}(u)$, then $f(u) < f^{**}(u)$.
\end{prop}

\noi 
{\bf Proof.} For any $u$ and $\beta$ in $\rsigma$ 
we have $f(u) \leq \lan \beta,u \ran - f^*(\beta)$ and thus
\[
f(u) \leq \inf_{\beta \in \rsigma}\{\lan \beta,u \ran - f^*(\beta)\} = f^{**}(u).
\]
If $\varphi$ is any concave, upper semicontinuous function satisfying
$\varphi(u) \geq f(u)$ for all $u$, then 
$\varphi^*(\beta) \leq f^*(\beta)$ for all $\beta$, and so 
$\varphi^{**}(u) = \varphi(u) \geq f^{**}(u)$
for all $u$.  \ \ink

\skp
Let $f \not \equiv -\infty$ be a function mapping $\rsigma$ into
$\r \cup \{-\infty\}$, $u$ a point in $\mboxdomf$,
and $K$ a convex subset of $\mboxdomf$.  Since $f^{**}$
is concave on $\rsigma$, the first three of the following
four definitions are consistent with Proposition \ref{prop:susu}:
$f$ is concave at $u$ if $f(u) = f^{**}(u)$; $f$ 
is not concave at $u$ if $f(u) < f^{**}(u)$;
$f$ is concave on $K$ if $f$ is concave at all $u \in K$;
and $f$ is strictly concave on $K$ if 
for all $u \not = v$ in $K$ and all $\lambda \in (0,1)$
\[
f(\lambda u + (1-\lambda)v) > \lambda f(u) + (1-\lambda) f(v).
\]

The next proposition gives a useful extension property of strictly concave functions.

\begin{prop}\per
\label{prop:extend}
Assume that $\mboxemdomf$ is convex and that $f$ is strictly concave on $\mboxemridomf$
and continuous on $\mboxemdomf$.  Then $f$ is concave on $\mboxemdomf$
and on $\rsigma$.
\end{prop}

\noi
{\bf Proof.}  Any point in $\mboxdomf \setminus \mboxintdomf$ is the limit
of a sequence of points in $\mboxridomf$ \cite[Thm.\ 6.1]{Rock}.
Hence by the continuity of $f$ on $\mboxdomf$, the strict concavity inequality 
for all $u \not = v$ in
$\mboxridomf$ can be extended to a nonstrict inequality for all $u$ and $v$ 
in $\mboxdomf$.  Hence $f$ is convex on $\mboxdomf$.  Since $f$ equals $-\infty$
on the complement of $\mboxdomf$, it also follows that $f$ is convex on $\rsigma$.
\ \ink

\skp
Parts (a) and (c) of the next theorem are fundamental in this paper because they relate concavity and support
properties of functions $f$ on $\rsigma$.  When applied to the microcanonical entropy
$s$ and to $s-g$, where $g$ is a continuous function defining the generalized 
canonical ensemble, part (c) of Theorem \ref{prop:2.9} allows us to deduce, from strict concavity properties
of $s$ and $s-g$, universal equivalence properties
involving the canonical ensemble and the generalized canonical ensemble.

\begin{thm}\per
\label{prop:2.9}
Let $f \not \equiv -\infty$ be a function mapping $\rsigma$ into $\r \cup \{-\infty\}$.
The following conclusions hold.

{\em (a)} 
$f$ has a supporting hyperplane at $u \in \mboxemdomf$ with normal vector $[\beta,-1]$
if and only if $f(u) = f^{**}(u)$ and $\beta \in \partial f^{**}(u)$.

{\em (b)} Assume that $f$ has a supporting hyperplane at $u \in \mboxemdomf$ 
with normal vector $[\beta,-1]$.  If $f$ is differentiable at $u$ and
$f = f^{**}$ in a neighborhood of $u$, then $\beta$ is unique
and $\beta = \nabla f(u)$.

{\em (c)} Assume that $\mboxemdomf$ is convex and that $f$ is strictly concave  
on $\mboxemridomf$ and continuous on $\mboxemdomf$.
Then $f$ has a strictly supporting hyperplane at all
$u \in \mboxemdomf$ except possibly relative boundary points.
In particular, if $\mboxemdomf$ is relatively open, then $f$ has a strictly supporting hyperplane at 
all $u \in \mboxemdomf$.
\end{thm}

\noi
{\bf Proof.}  
(a)  This is proved in part (a) of Lemma 4.1 in \cite{EHT1} when $f = s$.  The same
proof applies to general $f$.

(b) If $f$ has a supporting hyperplane at $u \in \mboxdomf$ with normal
vector $[\beta,-1]$, then by part (a), $\beta \in \partial f^{**}(u)$.  If
in addition $f$ is differentiable at $u$ and $f = f^{**}$ in a neighborhood
of $u$, then $f^{**}$ is also differentiable at $u$ and $\nabla f^{**}(u)
= \nabla f(u)$.  The conclusion that $\beta$ is unique and $\beta =
\nabla f(u)$ then follows from part (b) of Theorem \ref{thm:usefulfacts}
applied to $f^{**}$.

(c) By Proposition \ref{prop:extend} the assumptions on $f$ guarantee that 
$f$ is concave on $\rsigma$.  Since $\mboxridomf \subset \mbox{dom} \, \partial f$ 
[Thm.\ \ref{thm:usefulfacts}(a)], for any $u \in \mboxridomf$
and any $\beta \in \partial f(u)$, $f$ has a supporting hyperplane at $u$ 
with normal vector $[\beta,-1]$; i.e.,
\be
\label{eqn:supphyp2}
f(v) \leq f(u) + \lan \beta,v-u \ran \ \mbox{ for all } v \in \rsigma.
\ee
If this hyperplane is not a strictly supporting
hyperplane, then there exists $v_0 \not = u$ such that
\be
\label{eqn:getcontr1}
f(v_0) = f(u) + \lan \beta,v_0 - u \ran.
\ee
Thus $v_0 \in \mboxdomf$.  We claim that $f$ is strictly concave on 
$\mboxridomf \cup \{v_0\}$. If not, then $f$ must be 
affine on a line segment containing $v_0$.
Since this violates the strict concavity of $f$ on $\mboxridomf$, the
claim is proved.  Hence for all $\lambda \in (0,1)$
\[
\lambda f(u) + (1-\lambda) f(v_0) < f(\lambda u + (1-\lambda) v_0) \ \mbox{ for all }
\lambda \in (0,1).
\]
Substituting (\ref{eqn:getcontr1}) gives
\be
\label{eqn:getcontr2}
f(u) + (1 - \lambda) \lan \beta, v_0 - u \ran < f(\lambda u + (1-\lambda) v_0).
\ee
On the other hand, applying (\ref{eqn:supphyp2}) to $v = \lambda u + (1-\lambda)v_0$, 
we obtain
\beas
f(\lambda u + (1-\lambda)v_0) & \leq & f(u) + 
\lan \beta, \lambda u + (1-\lambda)v_0 - u \ran \\
& = & f(u) + (1-\lambda) \lan \beta, v_0 - u\ran.
\eeas  
This contradicts (\ref{eqn:getcontr2}), proving that the supporting hyperplane
at $u$ with normal vector $[\beta,-1]$ is a strictly supporting hyperplane.
We have proved that $f$ has a strictly supporting hyperplane at all $u \in \mboxridomf$
except possibly for relative boundary points.

If in addition $\mboxdomf$ is relatively open, then $\mboxridomf= \mboxdomf$.
It follows that in this case $f$ has a strictly supporting hyperplane at all
$u \in \mboxdomf$.  This completes the proof of part (b).  \ \ink

\skp
The next result is applied in Theorem \ref{thm:twolevels}, which relates ensemble
equivalence at the thermodynamic level and at the level of equilibrium 
macrostates.  Given $f \not \equiv -\infty$ a function
mapping $\rsigma$ into $\r \cup \{-\infty\}$, we define
\be
\label{eqn:C(f)}
C(f) = \{u \in \rsigma : \exists \beta \in \rsigma \ni f(v) \leq f(u) +
\lan \beta,v-u \ran \: \forall v \in \rsigma\}
\ee
and
\be
\label{eqn:Gamma}
\Gamma(f) = \{u \in \rsigma : f(u) = f^{**}(u)\}.
\ee
$C(f)$ consists of all $u \in \rsigma$ such 
that $f$ has a supporting hyperplane at $u$, and so if $u \in C(f)$,
then $\mbox{dom} \, \partial f(u) \not = \emptyset$.
In addition, $u \in \Gamma(f) \cap \mbox{dom} \, f$ if and only if $f$ is concave 
at $u$.  

\begin{thm}\per
\label{thm:5.7}
Let $f \not \equiv -\infty$ be a function mapping $\rsigma$ into 
$\r \cup \{-\infty\}$.  The following conclusions hold.

{\em (a)}  $C(f) = \Gamma(f) \cap \mbox{\em{dom}} \, \partial f^{**}$.
In particular, if $f$ is concave on $\rsigma$, then
$C(f) = \mbox{{\em dom}} \, \partial f$,
and so $f$ has a supporting hyperplane at all $u \in \mbox{{\em dom}} \, f$ 
except possibly relative boundary points.

{\em (b)} $\Gamma(f) \cap \mbox{\em{ri(dom}} \, f) \subset C(f) \subset \Gamma(f) \cap \mbox{\em{dom}} \, f$.

{\em (c)} Except possibly for relative boundary points of $\mboxemdomf$,
$f$ has no supporting hyperplane
at $u \in \mboxemdomf$ if and only if $f$ is not concave at $u$.
\end{thm}

\noi
{\bf Proof.} (a) The assertion that 
$C(f) = \Gamma(f) \cap \mbox{dom} \, \partial f^{**}$
is a consequence of part (a) of Theorem \ref{prop:2.9}.
Now assume that $f$ is concave on $\rsigma$.  Then, since
$f = f^{**}$, it follows that $\Gamma(f) = \rsigma$, $\mbox{dom} \, \partial f^{**}
= \mbox{dom} \, \partial f$, and thus $C(f) = \mbox{dom} \, \partial f$.  
Part (a) of Theorem \ref{thm:usefulfacts} implies that
$f$ has a supporting hyperplane at all points in $\mbox{dom} \, f$ 
except possibly relative boundary points.

(b) If $u \in \Gamma(f) \cap \mbox{ri(dom} \, f)$, 
then $f(u) = f^{**}(u)$ and $u \in \mbox{ri(dom} \, f^{**})$,
which in turn is a subset of $\mbox{dom} \, \partial f^{**}$ [Thm.\ \ref{thm:usefulfacts}(a)].
Hence $\Gamma(f) \cap \mbox{ri(dom} \, f) \subset \Gamma(f) \cap \mbox{dom} \, \partial f^{**}$,
which by part (a) equals $C(f)$.  This proves the first inclusion in part (b).  
To prove the second inclusion, we note that by part (a) $C(f) \subset \Gamma(f)$
and that for all $u \in C$, $f(u) > -\infty$.  Thus $C(f) \subset \Gamma(f) \cap \mboxdomf$.  

(c) If $f$ has no supporting hyperplane at $u \in \mboxridomf$, then 
$u \not \in C(f)$, and so by the first inclusion in part (b) $f \not \in \Gamma(f)$;
i.e., $f$ is not concave at $u$.  
Conversely, if $f$ is not concave at $u \in \mboxdomf$, then $u\not \in \Gamma(f)$, and so by 
the second inclusion in part (b) $u \not \in C(f)$; i.e.,
$f$ has no supporting hyperplane at $u$.  \ \ink

\newpage

\begin{acknowledgments}
The research of Marius Costeniuc and Richard S.\ Ellis
was supported by a grant from the National Science Foundation (NSF-DMS-0202309),
the research of Bruce Turkington was supported by a grant from the National Science
Foundation (NSF-DMS-0207064), and the research of Hugo Touchette was supported by
the Natural Sciences and Engineering Research Council of Canada and the Royal
Society of London (Canada-UK Millennium Fellowship).
\end{acknowledgments}

\end{document}